\documentclass[seceq,preprint]{ptptex}

\usepackage{graphicx}

\newcommand{\calc}[1]{\mathcal{{#1}}}
\newcommand{\lips}{d\textrm{LIPS}(l_2,-l_1;P_{L;z})}
\newcommand{\lipsP}{d\textrm{LIPS}(l_2,-l_1;P)}
\newcommand{\lr}[1]{\langle #1\rangle}
\newcommand{\denom}{\lr{(m_1-1)l_1}\lr{l_1m_1}\lr{m_2l_2}\lr{l_2(m_2+1)}}
\newcommand{\factor}{\frac{\lr{(m_1-1)m_1}\lr{m_2(m_2+1)}}{\denom}}

\newcommand{\tr}{\textrm{tr}}
\newcommand{\fsl}[1]{#1\hspace{-1.5mm}/}

\notypesetlogo                       
\preprintnumber[3cm]{
hep-th/0604210\\YITP-06-20\\
April, 2006
}

\title{
One-Loop Amplitudes in\\
Supersymmetric QCD from MHV Vertices
}

\author{
Hiroshi \textsc{Kunitomo}%
}

\inst{
Yukawa Institute for Theoretical Physics,\\
Kyoto University, Kyoto 606-8502, Japan
}

\abst{
The Cachazo-Svrcek-Witten (CSW) rule for efficiently calculating 
gauge theory amplitudes is extended to $\calc{N}=1$ supersymmetric
QCD (SQCD), incorporating massless quarks, in a manner that preserves the manifest 
supersymmetry. Using this extended CSW rule, 
we obtain compact expressions for all the one-loop MHV amplitudes 
in SQCD including one or two external quark-antiquark 
(chiral-antichiral multiplets) pairs. The collinear singularities 
of the five-point amplitudes are investigated to confirm 
the consistency of the results.
}

\begin{document}

\maketitle

\section{Introduction}\label{intro}

A new type of weak-weak duality proposed by Witten gives
a correspondence between the $\calc{N}=4$ super 
Yang-Mills (SYM) theory and topological strings on supertwistor 
space.\cite{W} The perturbative amplitudes
of the $\calc{N}=4$ SYM theory can be interpreted as 
D-instanton contributions in a topological string.

From the perturbative SYM viewpoint, Cachazo, Svrcek and
Witten showed that the essence of this duality can be extracted 
as a rule\cite{CSW} which provides an alternative
method to efficiently compute the perturbative YM amplitudes.
The building blocks of the CSW rule are tree-level maximally 
helicity violating (MHV) amplitudes and the scalar propagator, 
$1/P^2$. It also includes a definite prescription 
for using the (on-shell) MHV amplitudes as (off-shell) vertices. 
In comparison with the usual Feynman rule, 
the CSW rule has many practical advantages that simplify
the computation of amplitudes. 
The CSW rule was immediately extended to incorporate (extended)
supersymmetry\cite{GK}\tocite{BGK} 
and applied to obtain several gluon tree amplitudes.\cite{CSW,WZ} 
It was also applied to calculate the one-loop gluon MHV amplitudes 
of the $\calc{N}=4$\cite{BST} and $\calc{N}=1$\cite{BBST,QR} 
SYM theory, which reproduce previously obtained
results.\cite{BDDK4,BDDK1} 

Another extension of the CSW rule is obtained 
by introducing massless quarks
in the fundamental representation.\cite{WZ2,SW}\footnote{
See also Refs.~\citen{PR} and \citen{GKRRTZ}, in which
some tree-level MHV amplitudes involving quark-antiquark pairs 
in supersymmetric quiver gauge theories were studied.}
In this case, there are two types of new MHV amplitudes 
including one or two external quark-antiquark pairs, 
which must be added as extra vertices.
Several tree-level amplitudes with external fermions have
been computed by using this extended CSW rule 
and shown to be completely consistent with the results obtained 
using the usual Feynman rules.\cite{WZ2,SW}

The purpose of this paper is to combine these two
extensions. By introducing a scalar partner in the fundamental 
representation (squark), we derive an extended CSW rule 
for the $\calc{N}=1$ supersymmetric QCD (SQCD). 
We use a superfield formulation and adopt
the momentum representation of the supertwistor space 
by introducing some fermionic variables.\cite{BST}
Because this formulation involves only the physical helicity 
states without auxiliary fields, it provides a drastically 
simple alternative method efficiently compute the perturbative 
SQCD amplitudes, in a manner that preserves the manifest 
supersymmetry.
As an application of this extended CSW rule, we calculate 
all the one-loop MHV amplitudes with arbitrary numbers of 
legs including one or two external quark-antiquark pairs.
The supersymmetry acts non-trivially, because the superpartners 
contribute through the internal loop.
The new rule, using super-vertices, drastically reduces 
the number of Feynman-like diagrams, called MHV diagrams,
which we must sum up.
Only three (eight) MHV diagrams are needed
for the amplitudes, including one (two) external 
quark-antiquark pairs.
However, because we have no string theory interpretation,
there is no a priori reason that this extended CSW rule 
should lead to correct results in general. For this reason,
as a non-trivial check, we confirm that the five-point 
amplitudes, as examples, have the correct collinear singularities. 

This paper is organized as follows.
In \S\ref{treeMHV}, we rearrange the known tree-level MHV 
amplitudes in SQCD in a manifestly supersymmetric form
by introducing three fermionic variables, $\eta,\ \chi$ and $\rho$.
These supersymmetric MHV amplitudes are used as the MHV
vertices in our extended CSW rule, which is given 
at the beginning of \S\ref{oneloopamp}. 
Using this rule, we compute all the one-loop MHV amplitudes 
in $\calc{N}=1$ SQCD. In order to make this paper
self-contained, in \S\S\ref{gluons} we first present 
the gluon one-loop MHV amplitudes, which have already been 
obtained.\cite{BBST,QR}. Compact expressions of the one-loop 
MHV amplitudes with one or two external quark-antiquark pairs 
are given in \S\S\ref{twoquark} and \ref{fourquark}, 
respectively. As a non-trivial check of the results,
we investigate the collinear singularities of 
five-point amplitudes in \S\ref{fivepoint}.
Section \ref{discuss} is devoted to a summary and discussion. 
Finally, five appendices are provided. 
In Appendix~\ref{sconv}, our spinor conventions are summarized. 
Some useful formulas are also given there. Definitions and properties 
of the functions appearing in the one-loop amplitudes are given 
in Appendix~\ref{functions}. The formulas needed to evaluate 
the phase space integral are presented in Appendix~\ref{psint}.
Tree-splitting amplitudes and four-point amplitudes,
which are needed to study the collinear behavior of five-point
amplitudes, are summarized in Appendices \ref{treesplit} 
and \ref{fourpoint}, respectively.

\section{Tree-level MHV amplitudes in $\calc{N}=1$ SQCD}\label{treeMHV}

In this section we present the tree-level MHV amplitudes
of the $\calc{N}=1$ SQCD in a manifestly supersymmetric way
by introducing fermionic variables.
We employ the $U(N_c)$ gauge group for simplicity.

Let us first consider partial amplitudes of $n$ gluons
with fixed color ordering multiplied by the color factor 
\begin{equation}\label{cfallgluon}
C^{(0)}=\textrm{Tr}(T^{a_1}T^{a_2}\cdots T^{a_n}).
\end{equation}
Here $a_k\ (k=1,2,\cdots,n)$ is the color (adjoint) index of 
the $k$-th gluon. The $U(N_c)$ generators $T^a$ are
$N_c\times N_c$ hermitian matrices normalized so as to satisfy 
$\textrm{Tr}(T^aT^b)=\delta^{ab}/2$. 
The MHV helicity structure of the amplitudes is such
that two of the external gluons have negative helicities and 
the other $n-2$ have positive helicities.
We refer to the part other than the color factor
and the delta function yielding total momentum conservation,
$i(2\pi)^4\delta^4(\sum_ip_i)$, as the \lq\lq amplitude'' 
in this paper.

When a particle is massless, and hence $p^2=0$, the momentum 
vector $p_\mu$ in four dimensions can be conveniently rewritten 
in terms of the commutative spinor $\lambda_a$
[and its complex conjugate, $\bar{\lambda}_{\dot a}=(\lambda_a)^\dag$]
as\footnote{The spinor conventions used in this paper are summarized in
Appendix~\ref{sconv}.}
\begin{equation}
p_{a\dot{a}}=(\sigma_\mu)_{a\dot a}p^\mu=\lambda_a\bar{\lambda}_{\dot a}.
\end{equation}
The helicity amplitudes are functions of these commutative spinors 
and the helicities of the external particles,
$\lambda_{ia},\ \bar{\lambda}_{i\dot a}$ and $h_i$. 
The tree-level MHV amplitudes of $n$ gluons 
can thus be written in the holomorphic form
\begin{equation}\label{agym}
\calc{A}^{YM}_n=
\frac{\lr{ij}^4}{\lr{12}
\lr{23}\cdots\lr{n1}},
\end{equation}
where the $i$-th and $j$-th gluons are assumed to have 
negative helicity, and the shorthand notation 
$\lr{ij}=\lr{\lambda_i\lambda_j}=\lambda_{ia}\lambda_j^a$ 
is used. We use the convention that 
the momenta of all the external particles are outgoing.

\begin{figure}[htp]
\centerline{
 \includegraphics[width=5cm]{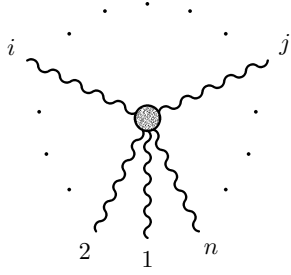}
}
   \caption{The $n$-gluon MHV vertex.}
    \label{fig:agmhv}
\end{figure}

In order to incorporate supersymmetry,
we also have to introduce two anticommuting variables,
$\eta$ and $\chi$, for $\calc{N}=1$ supersymmetry.\cite{W}
Using these variables, the helicity operator can be written
\begin{equation}\label{hel_vec}
h=1-\frac{1}{2}\eta\frac{\partial}{\partial\eta}
-\frac{3}{2}\chi\frac{\partial}{\partial\chi}.
\end{equation}
This is equivalent to the assumption that the vector superfield $V^a$ in
(the momentum representation of) the supertwistor space 
is a function of the variables $\lambda,\ \bar{\lambda},\ \eta$
and $\chi$ and can be expanded as
\begin{equation} \label{vecsf}
 V^a(\lambda,\bar{\lambda},\eta,\chi)=g^{a(+)}(\lambda,\bar{\lambda})
+\eta\Lambda^{a(+)}(\lambda,\bar{\lambda})
+\chi\Lambda^{a(-)}(\lambda,\bar{\lambda})
+\eta\chi g^{a(-)}(\lambda,\bar{\lambda}),
\end{equation}
where the component fields 
$g^{a(+)},\ \Lambda^{a(+)},\ \Lambda^{a(-)}$
and $g^{a(-)}$ has helicity $1,\ 1/2,\ -1/2$ and $-1$, respectively. 
It should be remarked that the fermionic variable
$\eta$ is directly related to the supersymmetry,
but $\chi$ is not. It has the role of combining the positive helicity 
multiplet $(g^{a(+)},\Lambda^{a(+)})$ and negative helicity multiplet
$(g^{a(-)},\Lambda^{a(-)})$ into one superfield.
Thus the tree-level MHV amplitudes of $n$ external $\calc{N}=1$
vector multiplets can be written as
\begin{align}\label{agtree}
 \calc{A}^{(0)}_n&=
-\sum_{k<l}(-1)^{\epsilon_k+\epsilon_l}\eta_k\eta_l
\sum_{i<j}(-1)^{\epsilon_i+\epsilon_j}\chi_i\chi_j
\frac{\lr{kl}\lr{ij}^3}{\lr{12}\lr{23}\cdots\lr{n1}}
\nonumber\\
&\equiv
\sum_{k<l}\sum_{i<j}\calc{A}^{(0)}_n(k,l,i,j),
\end{align}
where $\epsilon_i=0\ (1)$ if the momentum of the $i$-th particle 
is outgoing (incoming).\footnote{
In order to use them as vertices, we must also consider 
the incoming momenta, because the legs of a vertex do not 
have to be external.} 
We use these tree-level MHV amplitudes
as the vertices depicted in Fig.~\ref{fig:agmhv}.

The amplitude (\ref{agtree}) can be expanded in terms of
the anticommuting variables $\eta_i$ and $\chi_i$, 
and each coefficient corresponds to the amplitude of external 
states whose helicities can be read off by using the helicity 
operator (\ref{hel_vec}).
For example, the coefficient of
$\eta_i\chi_i\eta_j\chi_j$, with fixed $i$ and $j$,
is the $n$ gluon MHV amplitude (\ref{agym}).

Here we rewrite the amplitude (\ref{agtree}) as
a product of three parts for later use:
\begin{subequations}
\begin{align}
 \calc{A}^{(0)}_n&=\Delta(\eta)\Delta^{(0)}(\chi)
\prod_{i=1}^n\frac{1}{\lr{i(i+1)}},
\qquad (n+1\equiv 1)\\
\Delta(\eta)&=\sum_{k<l}(-1)^{\epsilon_k+\epsilon_l}
\lr{kl}\eta_k\eta_l=\delta^2\left(
\sum_k(-1)^{\epsilon_k}(\lambda^a_k\eta_k)\right),\\
\Delta^{(0)}(\chi)&=-\sum_{i<j}
(-1)^{\epsilon_i+\epsilon_j}\lr{ij}^3\chi_i\chi_j.
\end{align}
 \end{subequations}
It should be noted that
the first factor, $\Delta(\eta)$, can be written in the form 
of a delta function. The second factor, $\Delta^{(0)}(\chi)$,
however, does not have such an interpretation.
This is probably related to the unusual role of the $\chi$ variable 
mentioned above and makes it difficult to interpret it as amplitudes
of some string theory. 

In order to incorporate quark chiral multiplets,
we need to introduce another fermionic variable, $\rho$.
The quark chiral superfield $Q_i$ is a fermionic superfield and 
a function of the variables $\lambda,\ \bar{\lambda},\ \eta$
and $\rho$. It can be expanded as
\begin{equation} \label{chisf}
 Q_i(\lambda,\bar{\lambda},\eta,\rho)=q^{(+)}_i(\lambda,\bar{\lambda})
+\eta\phi^{(+)}_i(\lambda,\bar{\lambda})
+\rho\phi^{(-)}_i(\lambda,\bar{\lambda})
+\eta\rho q^{(-)}_i(\lambda,\bar{\lambda}).
\end{equation}
The component fields
$q^{(+)}_i,\ \phi^{(+)}_i,\ \phi^{(-)}_i$ and $q^{(-)}_i$
have the helicities $1/2,\ 0,\ 0$ and $-1/2$, respectively, 
and form a chiral multiplet of
four-dimensional $\calc{N}=1$ supersymmetry.
The helicity operator for the chiral superfield is given by
\begin{equation}\label{hel_chi}
h_c=\frac{1}{2}-\frac{1}{2}\eta\frac{\partial}{\partial\eta}
-\frac{1}{2}\rho\frac{\partial}{\partial\rho}.
\end{equation}
The new fermionic variable $\rho$ combines positive-helicity and 
negative-helicity multiplets, $(q^{(+)}_i,\phi^{(+)}_i)$ and 
$(\phi^{(-)}_i,q^{(-)}_i)$, into the single superfield (\ref{chisf}). 
The antiquark antichiral superfield $ \bar{Q}^i$ has properties
similar to those of the quark superfield, except for the color quantum
number. The (anti)quark superfield $Q_i$ ($\bar{Q}^i$) is in the 
(anti)fundamental representation of the gauge group $U(N_c)$.

\begin{figure}[htp]
\centerline{
 \includegraphics[width=10cm]{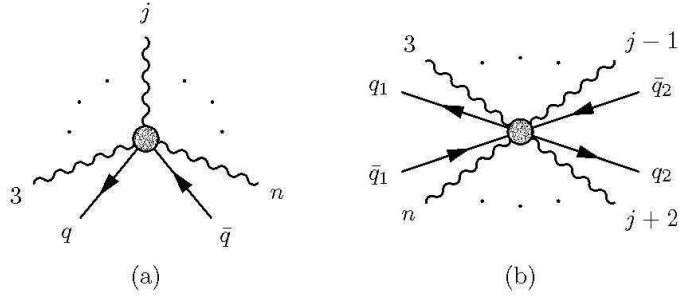}
}
   \caption{The MHV vertices including 
(a) one external quark-antiquark pair (b) two external quark-antiquark pairs.}
    \label{fig:qmhv}
\end{figure}

When quark multiplets are incorporated,
there are two types of additional MHV amplitudes, 
which we call two-quark and four-quark MHV amplitudes,
including one or two external quark-antiquark pairs
ordered as depicted in Fig.~\ref{fig:qmhv}.
We note that the fermion line, denoted by the solid line, 
is directed\footnote{
This should not be confused with the direction of the momentum.}
from the antiquark $\bar{q}$ to the quark $q$, and along this line,
the helicity is always conserved. The tree-level two-quark MHV amplitudes
$\calc{A}^{(2)}_n$,\cite{PRep} depicted in Fig.~\ref{fig:qmhv} (a), 
can be written in terms of the fermionic variables $\eta$ and $\rho$ as
\begin{align}\label{2qtree}
 \calc{A}^{(2)}_n&=
-\sum_{k<l=1}^n
(-1)^{\epsilon_k+\epsilon_l}\eta_k\eta_l
\sum_{\alpha=1}^2\sum_{j=3}^n
(-1)^{\epsilon_\alpha+\epsilon_j}\rho_\alpha\chi_j
\frac{\lr{kl}\lr{1j}\lr{2j}
\lr{\alpha j}} 
{\lr{12}\lr{23}\cdots\lr{n1}}
\nonumber\\
&\equiv
\sum_{k<l=1}^n\sum_{\alpha=1}^2\sum_{j=3}^n
\calc{A}^{(2)}_n(k,l,\alpha,j),
\end{align}
where $\epsilon_\alpha=0\ (1)$ if $\alpha=q\ (\bar{q})$, and
we also denote $\bar{q}$ by $1$ and $q$ by $2$.
The color factor,
\begin{equation}
C^{(2)}={(T^{a_3}\cdots T^{a_n})_{i_q}}^{i_{\bar q}},
\end{equation}
is omitted. This can also be written as a product:
\begin{subequations}
\begin{align}
\calc{A}^{(2)}_n&=\Delta(\eta)\Delta^{(2)}(\rho,\chi)
\prod_{i=1}^n\frac{1}{\lr{i(i+1)}},\\
\Delta^{(2)}(\rho,\chi)&=-\sum_{\alpha=1}^2\sum_{j=3}^n
(-1)^{\epsilon_\alpha+\epsilon_j}
\lr{1j}\lr{2j}\lr{\alpha j}\rho_\alpha\chi_j.
\end{align}
\end{subequations}

The tree-level four-quark MHV amplitudes
$\calc{A}^{(4)}_n(j)$ have the form
\begin{align}\label{4qtree}
\calc{A}^{(4)}_n(j)&=
-\sum_{k<l=1}^n
(-1)^{\epsilon_k+\epsilon_l}\eta_k\eta_l
\sum_{\alpha=1}^2\sum_{\beta=j}^{j+1}
(-1)^{\epsilon_\alpha+\epsilon_\beta}\rho_\alpha\rho_\beta
\frac{\lr{kl}
\lr{1(j+1)}\lr{2j}\lr{\alpha\beta}}
{\lr{12}\lr{23}\cdots\lr{n1}}
\nonumber\\
&\equiv
\sum_{k<l=1}^n
\sum_{\alpha=1}^2\sum_{\beta=j}^{j+1}
\calc{A}^{(4)}_n(k,l,\alpha,\beta),
\end{align}
where the color factor
\begin{equation}
C^{(4)}={(T^{a_3}\cdots T^{a_{j-1}})_{i_{q_1}}}^{i_{\bar{q}_2}}
{(T^{a_{j+2}}\cdots T^{a_n})_{i_{q_2}}}^{i_{\bar{q}_1}},
\end{equation}
is omitted and $\bar{q}_1,\ q_1,\ \bar{q}_2$ and $q_2$ are also
denoted by $1,\ 2,\ j$ and $j+1$, respectively,
for simplicity. 
The amplitudes are distinguished by $j=3,4,\cdots,n-1$, which
determine the manner in which
$n-4$ vector multiplets are divided into two sets,
as depicted in Fig.~\ref{fig:qmhv} (b). 
We can write the amplitude (\ref{4qtree}) in the product form
\begin{subequations}
\begin{align}
 \calc{A}^{(4)}_n(j)&=\Delta(\eta)\Delta^{(4)}(\rho)
\prod_{i=1}^n\frac{1}{\lr{i(i+1)}},\\
\Delta^{(4)}(\rho)&=-\sum_{\alpha=1}^2\sum_{\beta=j}^{j+1}
(-1)^{\epsilon_\alpha+\epsilon_\beta}
\lr{1(j+1)}\lr{2j}\lr{\alpha\beta}\rho_\alpha\rho_\beta.
\end{align}
\end{subequations}

\section{Extended CSW rule and one-loop MHV amplitudes in $\calc{N}=1$
SQCD}\label{oneloopamp}

Another important piece of the CSW rule is a prescription for
extending the external momenta of the tree-level MHV amplitudes
into off-shell momenta to take the amplitudes as the vertices
and to connect them by propagators.
The method appropriate for the one-loop calculation is to construct 
an off-shell momentum $L_\mu$ from the on-shell one $l_\mu,\ l^2=0$, 
and a real variable $z$ by using an arbitrary but fixed null vector 
$\hat{\eta}_\mu,\ \hat{\eta}^2=0$ 
as\cite{BST}\tocite{QR}\cite{BBST2,BSTnew}
\begin{equation}\label{offshellmom}
L_\mu=l_\mu+z\hat{\eta}_\mu.
\end{equation}

Including this prescription, the extended CSW rule 
for $\calc{N}=1$ SQCD is summarized as follows.

\begin{enumerate}
 \item Draw all the MHV diagrams with given external legs
that have the appropriate topology fixed by the color and helicity 
configurations.
 \item Assign the following building blocks for each diagram.
       \begin{enumerate}
	\item Vertex: the tree-level MHV amplitude (\ref{agtree}),
	      (\ref{2qtree}) or (\ref{4qtree}) with a delta function 
	      for total momentum conservation, represented by
$i(2\pi)^4\delta^4(\sum_ip_i)$.
	\item Propagator: $1/L^2$ using the off-shell momentum extended 
	      by the prescription (\ref{offshellmom}).
       \end{enumerate}
Two momenta and color indices connected by a propagator must be chosen
       so as to be conserved. 
 \item Assign the integral
$\int \frac{d^4L}{(2\pi)^4}d\eta d\psi$ to each propagator,
where we have $d\psi=d\chi$ ($d\rho$) for the propagator 
of a vector (chiral) multiplet.
An extra minus sign is also assigned to the chiral multiplet loop.
This comes from the fact that the quark superfield $Q$, given in 
(\ref{chisf}), is fermionic.
\end{enumerate}
We calculate all the one-loop MHV amplitudes
as an application of this extended CSW rule for $\calc{N}=1$ SQCD.

Before considering individual amplitudes,
let us first outline the general structure of 
the one-loop MHV amplitudes obtained from the CSW rule.
The MHV diagrams producing one-loop MHV amplitudes are depicted 
in Figs.~\ref{fig:ag1loop}-\ref{fig:4q1loop}.
All these diagrams have the same topology,
with two vertices connected by two propagators.
Each partial amplitude for a fixed diagram
is schematically given by\cite{BST}\tocite{BBST2}
\begin{equation}\label{mom1loop}
\calc{A}_n=
\int\frac{d^4L_1}{(2\pi)^4}
\frac{d^4L_2}{(2\pi)^4}
(2\pi)^4i
\delta^4(L_2-L_1+P_L)
d\eta_{l_1} d\psi_{l_1}d\eta_{l_2}d\psi_{l_2} 
\frac{1}{L_1^2}\calc{A}(L)\frac{1}{L_2^2}\calc{A}(R), 
\end{equation}
where $P_{L}$ is the sum of external momenta attached to the left 
vertex. The factor yielding total momentum conservation,
$i(2\pi)^4\delta^4(\sum_i p_i)$,
and the one-loop color factor $N_cC^{(0,2,4)}$ are omitted, as mentioned
above. The amplitude is obtained by summing up the
contributions from all the possible MHV diagrams. We use the
convention that all the external momenta are outgoing and loop momentum
flows in the clockwise direction.
The factor $\calc{A}(L) (\calc{A}(R))$ is an appropriate MHV vertex
corresponding to the left (right) vertex in the MHV diagram.

Using the variables ($l, z$)
in Eq.~(\ref{offshellmom}), the loop-integration measure can be
rewritten as
\begin{equation}
\frac{d^4L_1}{L^2_1}\frac{d^4L_2}{L^2_2}
\delta^4(L_2-L_1+P_L)=\frac{dz_1}{z_1}\frac{dz_2}{z_2}\lips,
\end{equation}
where the Lorentz-invariant phase-space integration measure
$\lips$ is defined by
\begin{equation}
\lips=d^4l_1\delta^{(+)}(l_1^2)d^4l_2\delta^{(+)}(l_2^2)
\delta^4(l_2-l_1+P_{L;z}),
\end{equation}
with $\delta^{(+)}(l^2)=\theta(l^0)\delta(l^2)$, 
$z=z_1-z_2$ and $P_{L;z}=P_L-z\hat{\eta}$.

Because the loop integral in (\ref{mom1loop}) generally has both
ultraviolet and infrared divergences, 
we use the supersymmetric regularization, \textit{i.e.}
the four-dimensional helicity scheme.\cite{BK} That is,
after carrying out all the spinor calculations in four dimensions,
the loop integral is evaluated in $4-2\epsilon$ dimensions.
After some manipulations, the general form of the one-loop 
MHV amplitudes (\ref{mom1loop}) can be written as
\begin{equation}
\calc{A}_n=
\int d\calc{M}
d\eta_{l_1}d\eta_{l_2}
d\psi_{l_1}d\psi_{l_2} 
\calc{A}(L)\calc{A}(R),
\end{equation}
where
\begin{equation}
 d\calc{M}=\frac{\left(\mu^2\right)^{\epsilon}}{(2\pi)^{3-2\epsilon}}
\frac{dz}{z}\lips.
\end{equation}
The renormalization scale $\mu$ is introduced,
and the phase-space integral must be evaluated 
in $4-2\epsilon$ dimensions. 
Moreover, the $\eta$ integration can be easily
carried out independently of the specific diagrams,
since it can be written in the form of the conservation,
as mentioned above:
\begin{align}
\int d\eta_{l_1}d\eta_{l_2}
\Delta(\eta_L)\Delta(\eta_R)=
\lr{l_1l_2}\sum_{i<j=1}^n
\lr{ij}\eta_i\eta_j=
\lr{l_1l_2}\Delta(\eta).
\end{align}
The remaining parts depend on the individual diagrams
and are computed in the following subsections.

\subsection{All-gluon MHV amplitudes}\label{gluons}

\begin{figure}[htp]
\centerline{
 \includegraphics[width=10cm]{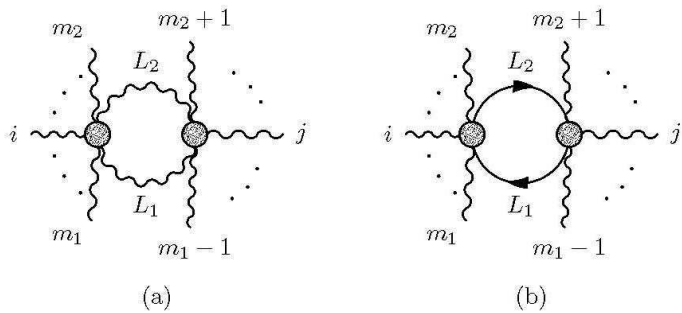}
}
   \caption{One-loop MHV diagrams for all-gluon amplitudes.}
    \label{fig:ag1loop}
\end{figure}

In order to specifically explain how the one-loop amplitudes 
are obtained from the CSW rule, we first consider the one-loop MHV 
amplitudes whose external particles are all gluon vector multiplets.
The superfield formulation naturally reproduces 
supersymmetrically decomposed amplitude (\ref{decom}).
The MHV diagrams contributing to such amplitudes are depicted
in Fig.~\ref{fig:ag1loop}. 
The first diagram, Fig.~\ref{fig:ag1loop} (a), represents 
the SYM contribution obtained in previous works.\cite{BBST,QR} 
The second one, Fig.~\ref{fig:ag1loop} (b), obtained by using
the new vertex $\calc{A}^{(2)}_n$, gives a new contribution coming
from the chiral-multiplet loop.

Let us first compute $\calc{A}^{(a)}_n$,
the contribution of the diagram (a). 
The loop $\chi$-integral can be easily evaluated,
after a calculation using Schouten's identity
(\ref{schouten}), as
\begin{align}
\int d\chi_{l_1}d\chi_{l_2}
\Delta^{(0)}(\chi_L)\Delta^{(0)}(\chi_R)=&
\lr{l_1l_2}^3
\sum_{i<j}\lr{ij}^3\chi_i\chi_j
\nonumber\\
&+3\lr{l_1l_2}
\sum_{i=m_1}^{m_2}\sum_{j=m_2+1}^{m_1-1}
\lr{ij}\lr{il_1}\lr{il_2}
\lr{jl_1}\lr{jl_2}\chi_i\chi_j.
\end{align}
Then we obtain
\begin{align}
\calc{A}^{(a)}_n=&
\sum_{k<l}\sum_{i<j}
\calc{A}^{(0)}_n(k,l,i,j)
\int d\calc{M} 
\frac{\lr{m_1(m_1-1)}\lr{l_2l_1}\lr{m_2(m_2+1)}\lr{l_1l_2}}
{\denom}
\nonumber\\
&-3\sum_{k<l}
\sum_{i=m_1}^{m_2}\sum_{j=m_2+1}^{m_1-1}
\calc{A}^{(0)}_n(k,l,i,j)
\int d\calc{M}
\frac{\lr{m_1(m_1-1)}\lr{m_2(m_2+1)}\lr{il_1}\lr{il_2}\lr{jl_1}\lr{jl_2}}
{\lr{ij}^2\denom}.
\end{align}
Each term here coincides with the contribution of 
the loop of the $\calc{N}=4$ vector and the $\calc{N}=1$ 
adjoint chiral multiplet obtained in Refs.~\citen{BST} 
and \citen{BBST,QR}, respectively. We have
\begin{equation}
 \calc{A}^{(a)}_n=\calc{A}^{\calc{N}=1 \textrm{vector}}_n=\calc{A}^{\calc{N}=4}_n
-3\calc{A}^{\calc{N}=1 \textrm{chiral}}_n,\label{decom}
\end{equation}
where the $\calc{N}=4$ SYM contribution
$\calc{A}^{\calc{N}=4}_n$ is given by\cite{BST}
\begin{equation}
\calc{A}^{\calc{N}=4}_n= 
\sum_{k<l}\sum_{i<j}
c_\Gamma\calc{A}^{(0)}_n(k,l,i,j)
\sum_{m=1}^n\sum_{r=1}^{\left[\frac{n}{2}\right]-1}
\left(1-\frac{1}{2}\delta_{\frac{n}{2}-1,r}\right)F(s,t,P^2,Q^2),
\end{equation}
with
\begin{equation}
 c_\Gamma=\frac{1}{(4\pi)^{2-\epsilon}}
\frac{\Gamma(1+\epsilon)\Gamma^2(1-\epsilon)}{\Gamma(1-2\epsilon)}.
\end{equation}
The summation is taken over all possible MHV diagrams 
of the type depicted in Fig.~\ref{fig:ag1loop} (a), which are obtained by
dividing the $n$ external legs into two vertices, fixing the ordering.
The factor of $1/2$ is needed for the diagrams whose two vertices 
have the same number, $n/2$, of legs because these are counted
twice in the summation.\cite{BST} The scalar box function $F$ and 
its finite part $B$ are defined by
\begin{align}
F(s,t,P^2,Q^2)=&-\frac{1}{\epsilon^2}
\left[
\left(\frac{\mu^2}{-s}\right)^\epsilon
+\left(\frac{\mu^2}{-t}\right)^\epsilon
-\left(\frac{\mu^2}{-P^2}\right)^\epsilon
-\left(\frac{\mu^2}{-Q^2}\right)^\epsilon
\right]
+B(s,t,P^2,Q^2),\\
B(s,t,P^2.Q^2)=&
\textrm{Li}_2\left(1-\frac{P^2}{s}\right)
+\textrm{Li}_2\left(1-\frac{P^2}{t}\right)
+\textrm{Li}_2\left(1-\frac{Q^2}{s}\right)
\nonumber\\
&\hspace{20mm}
+\textrm{Li}_2\left(1-\frac{Q^2}{t}\right)
-\textrm{Li}_2\left(1-\frac{P^2Q^2}{st}\right)
+\frac{1}{2}\log^2\left(\frac{s}{t}\right),
\label{box}
\end{align}
where $\textrm{Li}_2(x)=-\int^x_0\frac{\log(1-y)}{y}dy$ 
is the dilogarithm function\cite{dilog}.
The scalar invariants in the arguments 
are defined by $s=(p+P)^2$ and $ t=(P+q)^2$, with
$p=p_m,\ q=p_{m+r+1},\
P=p_{m+1}+\cdots+p_{m+r}$ and
$Q=p_{m+r+2}+\cdots+p_{m-1}$, which satisfy $p+q+P+Q=0$,
due to the conservation of total momentum.

The contribution of the $\calc{N}=1$ chiral multiplet,
$\calc{A}^{\calc{N}=1 \textrm{chiral}}_n$, 
has the form\cite{BBST,QR}
\begin{align}
\calc{A}^{\calc{N}=1 \textrm{chiral}}_n&= 
\sum_{k<l}\sum_{i<j}
c_\Gamma N_c\calc{A}^{(0)}_n(k,l,i,j)
\Bigg[
\sum_{m=j+1}^{i-1}\sum_{a=i+1}^{j-1}b^{ij}_{ma}
B(s,t,P^2,Q^2)
\nonumber\\
&\hspace{35mm}
+\frac{1}{1-2\epsilon}\left(
\sum_{m=i+1}^{j-1}\sum_{a=j}^{i-1}
c^{ij}_{ma}T_\epsilon(p,P^2,Q^2)
+(i\longleftrightarrow j)\right)
\Bigg],
\end{align}
with coefficients
\begin{subequations}
\begin{align}
b^{ij}_{ma}&=\frac{\tr(ijma)\tr(ijam)}{s_{ij}^2s_{ma}^2},\\
c^{ij}_{ma}&=\frac{1}{2}\left[
\frac{\tr(ijam)}{s_{ij}s_{m a}}
-\frac{\tr(ij(a+1)m)}{s_{ij}s_{ma+1}}\right]
\frac{\tr(ijmP_{m+1 a})-\tr(jimP_{m+1 a})}{s_{ij}}.
\end{align}
\end{subequations}
Here we introduce the notation $s_{ij}=(p_i+p_j)^2$,
$P_{m+1 a}=p_{m+1}+p_{m+2}+\cdots+p_a$ and
$ \tr(ijkl)=\tr(\fsl{p}_i\fsl{p}_j\fsl{p}_k\fsl{p}_l)$,
with the trace convention defined in Appendix~\ref{sconv}.
The $\epsilon$-dependent triangle function $T_\epsilon$ is 
defined by\cite{BBST}
\begin{equation}
T_\epsilon(p,P,Q)
=
\frac{\left(\mu^2\right)^{\epsilon}}{\epsilon}
\frac{(-P^2)^{-\epsilon}-(-Q^2)^{-\epsilon}}
{Q^2-P^2},\label{etriangle}
\end{equation}
where the renormalization point $\mu^2$ dependence is included.
This extended definition also contains the bubble function,
which has a divergent contribution, as the degenerate case,\cite{BBST}
as summarized in Appendix~\ref{functions}.
As in the case of the $\calc{N}=4$ part,
the summation is taken over all possible MHV diagrams 
constrained now such that the $i$-th and $j$-th external
legs, corresponding to the negative helicity states, 
are attached to the different vertices.\cite{BBST,QR}

The contribution of the diagram (b) can be similarly calculated
by using the the MHV vertex $\calc{A}^{(2)}$. 
As the loop $\rho$-integral gives
\begin{equation}
 \int d\rho_{l_1}d\rho_{l_2} \Delta^{(2)}(\rho_L,\chi_L)\Delta^{(2)}(\rho_R,\chi_R)
=\lr{l_1l_2}\sum_{i=m_1}^{m_2}\sum_{j=m_2+1}^{m_1-1}\lr{ij}\lr{il_1}\lr{il_2}
\lr{jl_1}\lr{jl_2}\chi_i\chi_j,
\end{equation}
the result becomes 
$\calc{A}^{(b)}_n=\calc{A}^{\calc{N}=1 \textrm{chiral}}_n$,
taking into account an extra minus sign.
If we have $N_f$ chiral multiplets (flavors), the total one-loop all-gluon
MHV amplitude is given by
\begin{equation}
 \calc{A}_n=\calc{A}^{\calc{N}=4}_n-3\calc{A}^{\calc{N}=1 \textrm{chiral}}_n
+\frac{N_f}{N_c}\calc{A}^{\calc{N}=1 \textrm{chiral}}_n.
\end{equation}
The factor of $1/N_c$ comes from the color factor.

\subsection{Two-quark MHV amplitudes}\label{twoquark}

\begin{figure}[htp]
\centerline{
 \includegraphics[width=15cm]{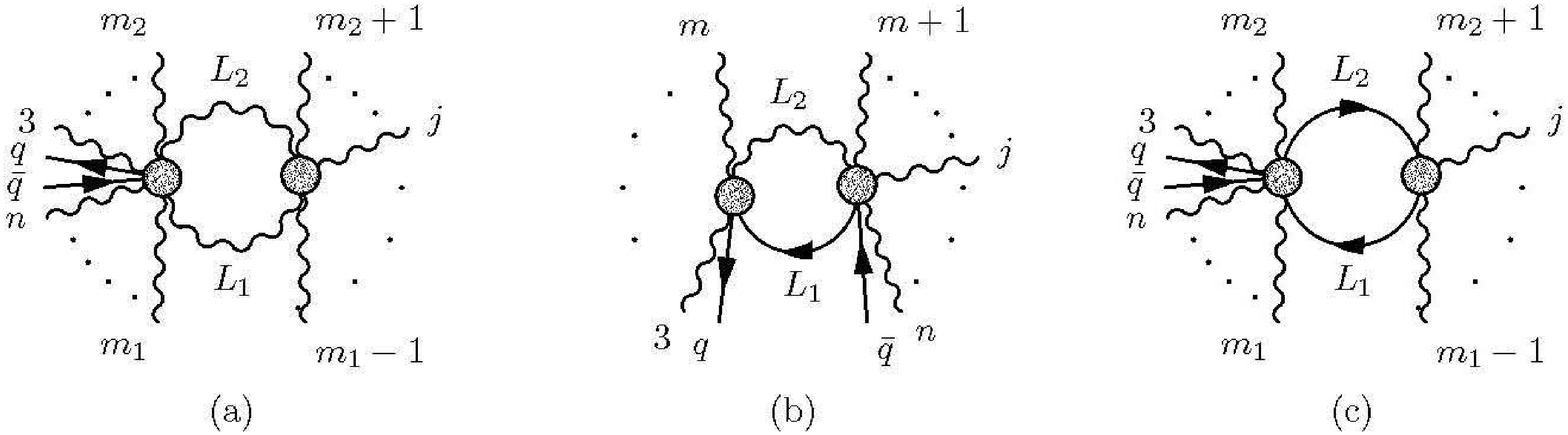}
}
   \caption{One-loop MHV diagrams for two-quark MHV amplitudes.}
    \label{fig:2q1loop}
\end{figure}

The one-loop two-quark MHV amplitudes
can be computed from the MHV diagrams depicted in Fig.~\ref{fig:2q1loop}.

Let us begin by evaluating the diagram (a).
The loop $\chi$-integral gives
\begin{align}\label{2quarka}
\int d\chi_{l_1}d\chi_{l_2}\Delta^{(2)}(L)
\Delta^{(0)}(R)
=&
\lr{l_1l_2}^3\sum_{\alpha=1}^2\sum_{j=3}^n\lr{1j}\lr{2j}
\lr{\alpha j}\rho_\alpha\chi_j
\nonumber\\
&+3\lr{l_1l_2}\sum_{\alpha=1}^2\sum_{j=m_2+1}^{m_1-1}
\lr{\check{\alpha}j}\lr{\alpha l_1}\lr{\alpha l_2}\lr{jl_1}\lr{jl_2}
\rho_\alpha\chi_j
\nonumber\\
&+\lr{l_1l_2}\sum_{\alpha=1}^2\sum_{j=m_2+1}^{m_1-1}
\lr{\alpha\check{\alpha}}\left(\lr{\alpha l_1}\lr{jl_2}+
\lr{\alpha l_2}\lr{jl_1}\right)\lr{jl_1}\lr{jl_2}\rho_\alpha\chi_j,
\end{align}
where $\check{\alpha}=2\ (1)$ for $\alpha=1\ (2)$.
The first two terms give, after summing up all the diagrams,
the same amplitudes $\calc{A}^{\calc{N}=4}_n$
and $\calc{A}^{\calc{N}=1,\textrm{chiral}}_n$,
as in the previous case. Actually, these amplitudes
can be decomposed into primitive amplitudes as\cite{BDK}
\begin{align}
\calc{A}_n&=
\calc{A}^{\mathcal{N}=4}_n-3\calc{A}^{\mathcal{N}=1 \textrm{chiral}}_n
-\calc{A}^R_n-\frac{N_f}{N_c}\calc{A}^f_n,
\label{dec2q}
\end{align}
where the last term is the contribution of the chiral-multiplet
loop coming from the diagram (c).
Therefore, here we compute the other contributions $\calc{A}^{R,f}_n$.
The contribution to $\calc{A}^R_n$ comes from the third term 
in Eq.~(\ref{2quarka}),
and yields
\begin{align}
\calc{A}^{R(a)}_n=& 
\sum_{k<l=1}^n\sum_{\alpha=1}^2\sum_{j=3}^n
\calc{A}^{(2)}_n(k,l,\alpha,j)
\sum_{m_1=j+1}^{n+1}\sum_{m_2=2}^{j-1}
\int d\calc{M}I^{(\alpha,j)(a)}_{(m_1,m_2)}(l_1,l_2),
\end{align}
where $n+1\equiv 1$ and
\begin{equation}
I^{(\alpha,j)(a)}_{(m_1,m_2)}(l_1,l_2)
=\frac{\lr{\alpha\check{\alpha}}\lr{(m_1-1)m_1}\lr{m_2(m_2+1)}
\left(\lr{\alpha l_1}\lr{jl_2}+\lr{\alpha l_2}\lr{jl_1}\right)\lr{jl_1}\lr{jl_2}}
{\lr{1j}\lr{2j}\lr{\alpha j}\denom}.
\end{equation}
The integrand $I^{(\alpha,j)(a)}_{(m_1,m_2)}(l_1,l_2)$ can be written as
\begin{align}
I^{(\alpha,j)(a)}_{(m_1,m_2)}(l_1,l_2)=&
R^{(\alpha,j)}(m_1-1,m_2)-R^{(\alpha,j)}(m_1-1,m_2+1)
\nonumber\\
&\hspace{20mm}
-R^{(\alpha,j)}(m_1,m_2)+R^{(\alpha,j)}(m_1,m_2+1), \label{4terms}
\end{align}
where
\begin{align}\label{integrand}
R^{(\alpha,j)}(m_1,m_2)
=&
\frac{1}{4(m_1\cdot l_1)(m_2\cdot l_2)}\Bigg(
\frac{\tr(j\alpha m_1l_1)\tr(j\check{\alpha}m_2l_2)}
{4(\alpha\cdot j)(\check{\alpha}\cdot j)}
+\frac{\tr(j\check{\alpha} m_1l_1)\tr(j\alpha m_2l_2)}
{4(\alpha\cdot j)(\check{\alpha}\cdot j)}
\nonumber\\
&\hspace{35mm}
-2\frac{\tr(j\alpha m_1l_1)\tr(j\alpha m_2l_2)}
{4(\alpha\cdot j)^2}
\Bigg)
\nonumber\\
\equiv&
\frac{r^{(\alpha,j)}_{(m_1,m_2)}(l_1,l_2)}
{4(m_1\cdot l_1)(m_2\cdot l_2)}.
\end{align}
We use the same notation for
(holomorphic) spinors and the corresponding four-vectors, 
e.g.
$\lr{ij}=\lr{\lambda_i\lambda_j}$ and
$(i\cdot j)=p_{i\mu}p_j^\mu$, with
$p_{ia\dot{a}}=\lambda_{ia}\bar{\lambda}_{i\dot{a}}$.

Next, we carry out the loop integration. Using the formula (\ref{twol}), 
the phase-space integral can be evaluated, except in the case 
$(m_1,m_2)=(1,2)$, which is considered later, as 
\begin{align}\label{int1}
\int d\calc{M}R^{(\alpha,j)}(m_1,m_2)=&
\frac{c_\Gamma\left(\mu^2\right)^{\epsilon}}{(\pi\epsilon\csc(\pi\epsilon))}
\int\frac{dz}{z}\left(P_{L;z}^2\right)^{-\epsilon}
\Bigg(-\frac{r^{(\alpha,j)}_{(m_1,m_2)}(m_2,m_1)}{4(m_1\cdot m_2)^2}
\log\left(1-aP_{L;z}^2\right)
\nonumber\\
&
+\frac{1}{1-2\epsilon}\left[
\frac{r^{(\alpha,j)}_{(m_1,m_2)}(P_{L;z},m_1)}{4(m_1\cdot m_2)
(m_1\cdot P_{L;z})}
+\frac{r^{(\alpha,j)}_{(m_1,m_2)}(m_2,P_{L;z})}{4(m_1\cdot m_2)
(m_2\cdot P_{L;z})}
\right]\Bigg),
\end{align}
where we use the $\hat{\eta}=p_{m_1}$ or $p_{m_2}$ gauge so that
\begin{equation}
 a=\frac{(m_1\cdot m_2)}{(m_1\cdot m_2)P_{L}^2-2(m_1\cdot P_{L})
(m_2\cdot P_{L})}.\label{aP}
\end{equation}
It should be noted that in the center-of-mass frame, in which
we have $P_{L;z}=P_{L;z}(1,\mathbf{0})$, the $P_{L;z}$ dependence 
in $[\ ]$ in the integrand cancels out.\cite{BBST2}
The dispersion integral therefore reduces to the two types of
integrations evaluated in Refs.~\citen{BST}-\citen{BBST},\citen{BBST2}:
\begin{subequations}
\begin{align}
&\int\frac{dz}{z}\left(P_{L;z}^2\right)^{-\epsilon}
\log\left(1-aP_{L;z}^2\right)
=\left(\pi\epsilon\csc(\pi\epsilon)\right)
\textrm{Li}_2\left(1-aP_L^2\right),\label{dis1}\\
&\int\frac{dz}{z}\left(P_{L;z}^2\right)^{-\epsilon}
=\left(\pi\epsilon\csc(\pi\epsilon)\right)
\frac{\left(-P_L^2\right)^{-\epsilon}}{\epsilon}.\label{dis2}
\end{align}
\end{subequations}
Consequently, the dispersion integral in (\ref{int1}) is 
calculated as
\begin{align}
\int d\calc{M}R^{(\alpha,j)}(m_1,m_2)=&
c_\Gamma
\Bigg(
-\frac{r^{(\alpha,j)}_{(m_1,m_2)}(m_2,m_1)}
{4(m_1\cdot m_2)^2}\textrm{Li}_2\left(1-aP_L^2\right)
\nonumber\\
&+\frac{1}{1-2\epsilon}\left[
\frac{r^{(\alpha,j)}_{(m_1,m_2)}(P_L,m_1)}
{4(m_1\cdot m_2)(m_1\cdot P_L)}
+\frac{r^{(\alpha,j)}_{(m_1,m_2)}(m_2,P_L)}
{4(m_1\cdot m_2)(m_2\cdot P_L)}
\right]\frac{1}{\epsilon}
\left(\frac{\mu^2}{-P_L^2}\right)^{\epsilon}
\Bigg).\label{2quarkdis}
\end{align}
We must evaluate the integral separately
for the case $(m_1,m_2)=(1,2)$, because in that case,
$a$ defined by Eq.~(\ref{aP}) becomes infinity.
This appears only for the case $P_L=P_{12}$ and 
can be computed  by using the formula (\ref{onetwo})
as\footnote{This is effectively obtained by replacing the
corresponding (divergent) dilogarithm function $\textrm{Li}_2$ 
in (\ref{2quarkdis}) with the factor 
$-\frac{1}{\epsilon^2}\left(\frac{\mu^2}{-s_{12}}\right)^\epsilon$.
}
\begin{equation}
\int d\calc{M}|_{P_L=P_{12}}R^{(\alpha,j)}(1,2)=
\frac{c_\Gamma}{\epsilon^2(1-2\epsilon)}
\left(\frac{\mu^2}{-s_{12}}\right)^\epsilon.
\end{equation}

By adding up the four (two) terms in (\ref{4terms}) and  
appropriately shifting the summation indices $m_1$ and $m_2$, 
we generically get 
the box (triangle) function defined by (\ref{box}) 
((\ref{etriangle})) due to the formula (\ref{boxformula}).
However, some terms at the boundary of the summation
region do not have partners.

The missing pieces of the primitive amplitude
$\calc{A}^R_n$ are supplied by the diagram (b).
The loop fermion integration 
$\int d\chi_{l_1}d\rho_{l_2}\Delta^{(2)}(L)\Delta^{(2)}(R)$ 
in this case can be computed similarly, and the contribution
to the primitive amplitudes $\calc{A}^R_n$ is given by
\begin{align}
&
\lr{l_1l_2}\sum_{j=3}^{m_2}
\Big(\lr{2j}\lr{1l_1}\lr{jl_2}
\left(
\lr{1l_1}\lr{jl_2}+\lr{1l_2}\lr{jl_1}\right)\rho_1\chi_j
-\lr{2j}\lr{1l_1}\lr{jl_2}
\lr{l_1l_2}\lr{2j}\rho_2\chi_j\Big)
\nonumber\\
&+\lr{l_1l_2}
\sum_{j=m_2+1}^n\Big(
\lr{2j}\lr{1l_1}\lr{jl_2}
\lr{l_1l_2}\lr{1j}\rho_1\chi_j
-\lr{2j}\lr{1l_1}\lr{jl_2}
\left(
\lr{2l_1}\lr{jl_2}+\lr{2l_2}\lr{jl_1}\right)\rho_2\chi_j
\Big).
\end{align}
The amplitudes $\calc{A}^{R(b)}_n$ of the diagram (b) 
then become
\begin{align}
\calc{A}^{R(b)}_n=& 
\sum_{k<l=1}^n\sum_{\alpha=1}^2\sum_{j=3}^n
\calc{A}^{(2)}_n(k,l,\alpha,j)
\sum_{m=2}^{j-1}
\int d\calc{M}I^{(\alpha,j)(b)}_{(R,m)}(l_1,l_2)
\nonumber\\
&
+\sum_{k<l=1}^n\sum_{\alpha=1}^2\sum_{j=3}^n
\calc{A}^{(2)}_n(k,l,\alpha,j)
\sum_{m=j}^n
\int d\calc{M}I^{(\alpha,j)(b)}_{(L,m)}(l_1,l_2),
\end{align}
where
\begin{equation}
I^{(\alpha,j)(b)}_{(J,m)}
=\left\{
\begin{split}
&\frac{\lr{12}\lr{m(m+1)}
\left(\lr{\alpha l_1}\lr{jl_2}+\lr{jl_1}\lr{\alpha l_2}\right)\lr{jl_2}}
{\lr{\alpha j}^2\lr{\check{\alpha}l_1}\lr{ml_2}\lr{l_2(m+1)}},
&\textrm{for $(J,\alpha)=(L,1), (R,2)$},\\
&\frac{\lr{12}\lr{m(m+1)}\lr{l_1l_2}\lr{jl_2}}
{\lr{\check{\alpha}j}\lr{\alpha l_1}\lr{ml_2}\lr{l_2(m+1)}},
&\textrm{for $(J,\alpha)=(L,2), (R,1)$}.
\end{split}
\right.
\end{equation}
The integrand $I^{(\alpha,j){b}}_{(J,m)}$ can be rewritten as
\begin{equation}
 I^{(\alpha,j){b}}_{(J,m)}=R^{(\alpha,j)}_J(m)-R^{(\alpha,j)}_J(m+1),
\end{equation}
with
\begin{equation}
 R^{(\alpha,j)}_J(m)=
\left\{
\begin{split}
&\gamma(\alpha)\Bigg[\frac{1}{4(\check{\alpha}\cdot l_1)(m\cdot l_2)}
\left(
\frac{\tr(j\alpha\check{\alpha}l_1)\tr(j\check{\alpha}ml_2)}
{4(\alpha\cdot j)(\check{\alpha}\cdot j)}
-2\frac{\tr(j\alpha\check{\alpha}l_1)\tr(j\alpha ml_2)}
{4(\alpha\cdot j)^2}
\right)\\
&\hspace{6cm}
+
\frac{1}{2(m\cdot l_2)}\left(
\frac{\tr(j\alpha ml_2)}{2(\alpha \cdot j)}
-\frac{\tr(j\check{\alpha}ml_2)}{2(\check{\alpha}\cdot j)}\right)
\Bigg],
\\
&\hspace{9cm}
\textrm{for $(J,\alpha)=(L,1), (R,2)$},\\
&\gamma(\alpha)\Bigg[
-\frac{1}{4(\alpha\cdot l_1)(m\cdot l_2)}\left(
\frac{\tr(j\check{\alpha}\alpha l_1)
\tr(j\alpha ml_2)}{4(\alpha\cdot j)(\check{\alpha}\cdot j)}\right)
\\
&\hspace{6cm}
+\frac{1}{2(m\cdot l_2)}\left(
\frac{\tr(j\alpha ml_2)}{2(\alpha\cdot j)}
-\frac{\tr(j\check{\alpha}ml_2)}{2(\check{\alpha}\cdot j)}\right)
\Bigg],
\\
&
\hspace{9cm}
\textrm{for $(J,\alpha)=(L,2), (R,1)$},
\end{split}
\right.
\end{equation}
where $\gamma$ is a sign factor defined by
\begin{equation}
 \gamma(\alpha)=
\begin{cases}
 -1 &\textrm{for $\alpha=1$},\\
 +1 &\textrm{for $\alpha=2$}.
\end{cases}
\end{equation}
We can decompose $R^{(\alpha,j)}_J(m)$ as
\begin{equation}
 R^{(\alpha,j)}_J(m)=
\left\{
\begin{split}
&\gamma(\alpha)\left(
R^{(\alpha,j)}(\check{\alpha},m)+
\tilde{R}^{(\alpha,j)}(m)
\right),
&\textrm{for $(J,\alpha)=(L,1), (R,2)$},\\
&\gamma(\alpha)\left(
-R^{(\alpha,j)}(\alpha,m)+
\tilde{R}^{(\alpha,j)}(m)
\right),
&\textrm{for $(J,\alpha)=(L,2), (R,1)$},
\end{split}
\right.
\end{equation}
where
\begin{align}
\tilde{R}^{(\alpha,j)}(m)=&\frac{1}{2(m\cdot l_2)}\left(
\frac{\tr(j\alpha ml_2)}{2(\alpha \cdot j)}
-\frac{\tr(j\check{\alpha}ml_2)}{2(\check{\alpha}\cdot j)}\right)
\nonumber\\
=&
\frac{\tilde{r}^{(\alpha,j)}_m(l_2)}{2(m\cdot l_2)}.
\end{align}
The loop integral of the first term is given in
(\ref{int1}). The second term can be similarly computed as
\begin{align}
\int d\calc{M}\tilde{R}^{(\alpha,j)}(m)=&
\frac{c_\Gamma\left(\mu^2\right)^{\epsilon}}{
(\pi\epsilon\csc(\pi\epsilon))
(1-2\epsilon)
}
\int\frac{dz}{z}
\left(P_{L;z}^2\right)^{-\epsilon}
\frac{\tilde{r}^{(\alpha,j)}_m(P_{L;z})}{2(m\cdot P_{L;z})}
\nonumber\\
=&
\frac{c_\Gamma}{1-2\epsilon}
\frac{\tilde{r}^{(\alpha,j)}_m(P_{L})}{2(m\cdot P_{L})}
\frac{1}{\epsilon}
\left(\frac{\mu^2}{-P_L^2}\right)^{\epsilon},
\end{align}
using formulas (\ref{onel}) and (\ref{dis2}).
We replace $P_{L;z}$ with $P_L$, except for the 
$(P_{L;z}^2)^{-\epsilon}$ factor, in the dispersion integral,
for the same reason as above.

By summing up the contributions from the two diagrams (a) and (b),
the primitive amplitudes $\calc{A}^R_n$ of the two-quark MHV amplitudes
are, after some calculation, obtained as
\begin{align}\label{2quarkR}
\calc{A}^{R}_n=& 
\sum_{k<l=1}^n\sum_{\alpha=1}^2\sum_{j=3}^n
c_\Gamma\calc{A}^{(2)}_n(k,l,\alpha,j)
\Bigg[
-\frac{1}{\epsilon^2}\left(\frac{\mu^2}{-s_{12}}\right)^\epsilon
+{
\sum_{m=j+1}^{n+1}\sum_{\substack{a=2\\
\hspace{-12mm}(m,a)\ne(1,2)}}^{j-1}}b^{\alpha j}_{ma}B(m,a)
\nonumber\\
&
+\frac{1}{1-2\epsilon}
\left(
\sum_{m=2}^{j-1}\sum_{a=j}^n
+\sum_{m=j+1}^{n+1}\sum_{a=2}^{j-1}
\right)c^{\alpha j}_{ma}
T_\epsilon(m,a)
+\frac{1}{1-2\epsilon}
\sum_{m\in\Gamma(\alpha)}\tilde{c}^{\alpha j}_m
T_\epsilon(m,1)
\Bigg],
\end{align}
with
\begin{subequations} 
\begin{align}
b^{\alpha j}_{ma}=&
\frac{\tr(j\alpha ma)\tr(j\check{\alpha}am)}
{s_{j\alpha}s_{j\check{\alpha}}s_{ma}^2}
+\frac{\tr(j\check{\alpha}ma)\tr(j\alpha am)}
{s_{j\alpha}s_{j\check{\alpha}}s_{ma}^2}
-2\frac{\tr(j\alpha ma)\tr(j\alpha am)}
{s_{j\alpha}^2s_{ma}^2},\\
c^{\alpha j}_{ma}=&
\left(
\frac{\tr(j\alpha am)}{s_{j\alpha}s_{ma}}
-\frac{\tr(j\alpha(a+1)m)}{s_{j\alpha}s_{ma+1}}
\right)
\frac{\tr(j\check{\alpha}mP_{m+1a})}{s_{j\check{\alpha}}}
\nonumber\\
&+
\left(
\frac{\tr(j\check{\alpha}am)}{s_{j\check{\alpha}}s_{ma}}
-\frac{\tr(j\check{\alpha}(a+1)m)}{s_{j\check{\alpha}}s_{ma+1}}
\right)
\frac{\tr(j\alpha mP_{m+1a})}{s_{j\alpha}}
\nonumber\\
&-2
\left(
\frac{\tr(j\alpha am)}{s_{j\alpha}s_{ma}}
-
\frac{\tr(j\alpha(a+1)m)}{s_{j\alpha}s_{ma+1}}
\right)
\frac{\tr(j\alpha mP_{m+1a})}{s_{j\alpha}},\\
\tilde{c}^{\alpha j}_m
=&-2\gamma(\alpha)
\frac{\tr(j\alpha\check{\alpha}m)\tr(j\alpha mP_{m+11})}
{s_{\alpha j}^2s_{\check{\alpha}m}}, 
\end{align}
\end{subequations}
where the summation region in the last term is defined by
\begin{equation}
\Gamma(\alpha)=
\begin{cases}
 \{j+1,\cdots,n\}, &\textrm{for $\alpha=1$},\\
\{3,\cdots,j-1\}, &\textrm{for $\alpha=2$},
\end{cases} 
\end{equation}
and we have introduced the notation
\begin{subequations}
\begin{align}
 B(m,a)\equiv&B(P^2_{m a-1},P^2_{m+1 a},P^2_{m+1 a-1},P^2_{a+1 m-1}),\\
T_\epsilon(m,a)\equiv&T_\epsilon(p_m,P_{a+1m-1},P_{m+1a}).\label{triabb}
\end{align}
\end{subequations}

The quark-loop diagram (c) leads the primitive amplitude
$\calc{A}^f_n$ in the decomposition (\ref{dec2q}).
After carrying out the loop fermion integration, we have
\begin{align}
\calc{A}^f_n&=
 -\sum_{k<l=1}^n\sum_{\alpha=1}^2\sum_{j=3}^n
\calc{A}^{(2)}_n(k,l,\alpha,j)
\sum_{m_1=j+1}^{n+1}\sum_{m_2=2}^{j-1}
\int d\calc{M}I^{(j)}_{(m_1,m_2)}(l_1,l_2),
\end{align}
where
\begin{equation}
I^{(j)}_{(m_1,m_2)}(l_1,l_2)=
\frac{\lr{(m_1-1)m_1}\lr{m_2(m_2+1)}\lr{1l_1}\lr{2l_2}\lr{jl_1}\lr{jl_2}}
{\lr{1j}\lr{2j}\denom}.
\end{equation}
This can also be written as
\begin{equation}
 I^{(j)}_{(m_1,m_2)}(l_1,l_2)=
R_f^{(j)}(m_1-1,m_2)
-R_f^{(j)}(m_1-1,m_2+1)
-R_f^{(j)}(m_1,m_2)
+R_f^{(j)}(m_1,m_2+1),
\end{equation}
where
\begin{equation}
 R_f^{(j)}(m_1,m_2)=
-\frac{\tr(j1m_1k)\tr(j2m_2l)}
{16(1\cdot j)(2\cdot j)(m_1\cdot l_1)(m_2\cdot l_2)}.
\end{equation}
The loop integral can be
easily carried out similarly to those in the previous cases,
and we obtain
\begin{align}\label{2quarkf}
\calc{A}_n^f=&
\sum_{k<l=1}^n\sum_{\alpha=1}^2\sum_{j=3}^n
c_\Gamma\calc{A}^{(2)}_n(k,l,\alpha,j)\Bigg[
\sum_{m=j+1}^n\sum_{a=3}^{j-1}b^j_{ma}B(m,a)
\nonumber\\
&\hspace{20mm}
+\frac{1}{1-2\epsilon}\left(
\sum_{m=3}^{j-1}\sum_{a=j}^n
+\sum_{m=j+1}^n\sum_{a=2}^{j-1}
\right)
c^j_{ma}T_\epsilon(m,a)
\Bigg],
\end{align}
where the coefficient functions are defined by
\begin{subequations} 
\begin{align}
b^j_{ma}=&
-\frac{\tr(j1ma)\tr(j2am)}{s_{j1}s_{j2}s_{ma}^2},\\ 
c^j_{ma}=&
\frac{1}{2}\Bigg[\left(
\frac{\tr(1jam)}{s_{1j}s_{ma}}
-\frac{\tr(1j(a+1)m)}{s_{1j}s_{ma+1}}
\right)
\frac{\tr(j2mP_{m+1a})}{s_{2j}}
\nonumber\\
&\hspace{10mm}
+
\left(
\frac{\tr(2jam)}{s_{2j}s_{ma}}
-\frac{\tr(2j(a+1)m)}{s_{2j}s_{ma+1}}
\right)
\frac{\tr(j1mP_{m+1a})}{s_{1j}}
\Bigg].
\end{align}
\end{subequations}

\subsection{Four-quark MHV amplitudes}\label{fourquark}

\begin{figure}[htp]
\centerline{
 \includegraphics[width=15cm]{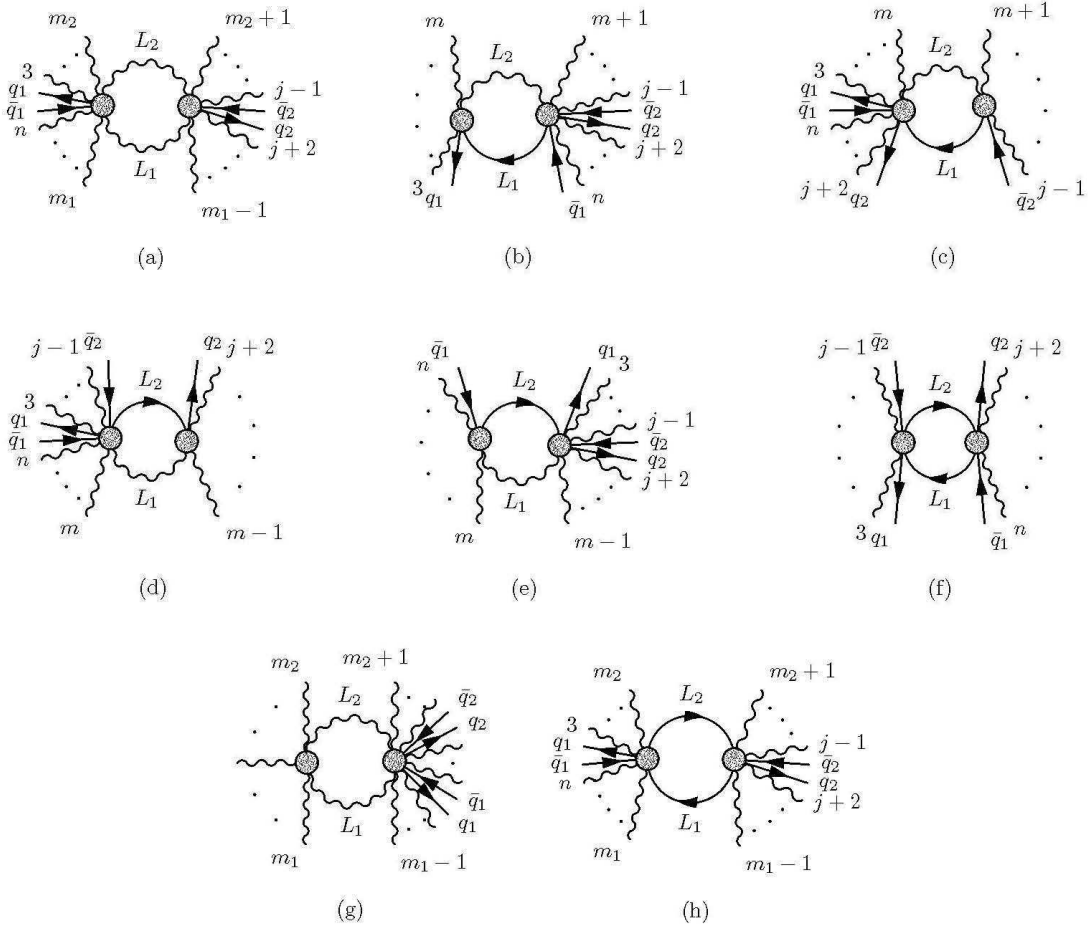}
}
   \caption{One-loop MHV diagrams for four-quark MHV amplitudes.}
    \label{fig:4q1loop}
\end{figure}

We can similarly calculate the one-loop four-quark MHV amplitudes 
from the MHV diagrams depicted Fig.~\ref{fig:4q1loop}. 
The seven diagrams (a)-(g)
give contributions to the first three terms of the same
decomposition as in the previous case, (\ref{dec2q}):
\begin{align}
\calc{A}_n&=
\calc{A}^{\mathcal{N}=4}_n-3\calc{A}^{\mathcal{N}=1 \textrm{chiral}}_n
-\calc{A}^R_n-\frac{N_f}{N_c}\calc{A}^f_n.
\nonumber
\end{align}
We present a compact expression of the unknown
piece $\calc{A}^R_n$ in the following. 

There are two types of amplitudes characterized
by the helicity configurations $(\alpha,\beta)=(1,j),\ (2,j+1)$
and $(1,j+1),\ (2,j)$. The two quarks have the same helicity
in the former configuration and different helicities in the latter.
After the loop fermion integration, the contribution
to $\calc{A}^{R(a)}_n$ from the diagram (a) becomes
\begin{equation}
\calc{A}^{R(a)}_n=
\sum_{k<l=1}^n\sum_{\alpha=1}^2\sum_{\beta=j}^{j+1}
\calc{A}^{(4)}_n(k,l,\alpha,\beta)
\sum_{m_1=j+2}^{n+1}\sum_{m_2=2}^{j-1}
\int d\calc{M}J^{(\alpha,\beta)(a)}_{(m_1,m_2)}(l_1,l_2),
\label{summation}
\end{equation}
where we have
\begin{equation}
 J^{(\alpha,\beta)(a)}_{(m_1,m_2)}(l_1,l_2)=
\left\{
\begin{split}
 &\Bigg(
-\frac{\lr{\alpha\check{\alpha}}\lr{\beta\check{\beta}}
\lr{\alpha l_1}\lr{\alpha l_2}\lr{\beta l_1}\lr{\beta l_2}}
{\lr{\alpha\check{\beta}}\lr{\check{\alpha}\beta}\lr{\alpha\beta}^2}\\
&\hspace{1cm}
+\frac{\lr{\alpha\check{\alpha}}\lr{\beta l_1}\lr{\beta l_2}
\left(
\lr{\alpha l_1}\lr{\beta l_2}+\lr{\beta l_1}\lr{\alpha l_2}
\right)}{\lr{\check{\alpha}\beta}\lr{\alpha\beta}^2}\\
&\hspace{2cm}
-\frac{\lr{\beta\check{\beta}}\lr{\alpha l_1}\lr{\alpha l_2}
\left(
\lr{\alpha l_1}\lr{\beta l_2}+\lr{\beta l_1}\lr{\alpha l_2}
\right)}{\lr{\alpha\check{\beta}}\lr{\alpha\beta}^2}
\Bigg)\calc{D}(m_1,m_2),\\
 &\hspace{5cm}
\textrm{for $(\alpha,\beta)=(1,j)$ or $(2,j+1)$},\\
 &\Bigg(
-2\frac{\lr{\alpha l_1}\lr{\alpha l_2}\lr{\beta l_1}\lr{\beta l_2}}
{\lr{\alpha\beta}^2}
\\
&\hspace{1cm}
+\frac{\lr{\alpha l_1}\lr{\check{\alpha}l_2}\lr{\check{\beta}l_1}\lr{\beta l_2}}
{\check{\alpha}\check{\beta}\lr{\alpha\beta}}
+\frac{\lr{\check{\alpha}l_1}\lr{\alpha l_2}\lr{\beta l_1}\lr{\check{\beta}l_2}}
{\check{\alpha}\check{\beta}\lr{\alpha\beta}}
\Bigg)\calc{D}(m_1,m_2),
\\
 &\hspace{5cm}
\textrm{for $(\alpha,\beta)=(1,j+1)$ or $(2,j)$}.
\end{split}
\right.
\end{equation}
Here we define the common factor $D$ as
\begin{equation}
 \calc{D}(m_1,m_2)=\factor.
\end{equation}
The integrand $J^{(\alpha,\beta)(a)}_{(m_1,m_2)}(l_1,l_2)$ can be
decomposed as
\begin{align}
 J^{(\alpha,\beta)(a)}_{(m_1,m_2)}(l_1,l_2)=&
S^{(\alpha,\beta)}(m_1-1,m_2)
-S^{(\alpha,\beta)}(m_1-1,m_2+1)
\nonumber\\
&\hspace{20mm}
-S^{(\alpha,\beta)}(m_1,m_2)
+S^{(\alpha,\beta)}(m_1,m_2+1),\label{jadec}
\end{align}
with
\begin{equation}
S^{(\alpha,\beta)}(m_1,m_2)
=
\left\{
\begin{split}
&
\frac{1}{4}(\Omega-4)
\left(
\calc{S}(\alpha,\beta,\alpha,\beta)
+\calc{S}(\beta,\alpha,\beta,\alpha)
\right)\\
&\hspace{1cm}
+\calc{S}(\alpha,\beta,\alpha,\check{\beta})
+\calc{S}(\beta,\alpha,\beta,\check{\alpha}),
\\
&\hspace{3cm}
\textrm{for $(\alpha,\beta)=(1,j)$ or $(2,j+1)$},\\
&
-\frac{1}{2}\left(\calc{S}(\alpha,\beta,\alpha,\beta)
+\calc{S}(\beta,\alpha,\beta,\alpha)\right)\\
&\hspace{1cm}
+\frac{1}{2}\Omega\left(
\calc{S}(\check{\alpha},\alpha,\check{\alpha},\beta)
+\calc{S}(\check{\beta},\beta,\check{\beta},\alpha)
\right)\\
&\hspace{2cm}
+\frac{1}{2}\left(
\calc{S}(\alpha,\beta,\alpha,\check{\beta})
+\calc{S}(\beta,\alpha,\beta,\check{\alpha})\right),
\\
&\hspace{3cm}
\textrm{for $(\alpha,\beta)=(1,j+1)$ or $(2,j)$},
\end{split}
\right.
\\
\end{equation}
where
\begin{align}
\Omega=&\frac{\tr(1(j+1)j2)}{4(1\cdot (j+1))(2\cdot j)},\\
\calc{S}(a,b,c,d)=&
\frac{\tr(abm_1l_1)\tr(cdm_2l_2)+\tr(cdm_1l_1)\tr(abm_2l_2)}
{16(a\cdot b)(c\cdot d)(m_1\cdot l_1)(m_2\cdot l_2)}.
\end{align}

The loop integral can be carried out similarly to 
those in the previous cases. 
We can sum the generic terms of (\ref{jadec}) 
in the summation (\ref{summation}) by renaming the indices,
and we thereby obtain some combinations of the box and triangle functions.
There are, however, some missing terms, for which the contributions
of the other diagrams, (b)-(g), compensate.

The four diagrams (b)-(e) have similar structures, and for this reason
it is sufficient to evaluate just two of them (b) and (c).
The primitive amplitudes $\calc{A}^{R(b,c)}_n$ after the fermion 
integration can be written as
\begin{equation}
\calc{A}^{R(b,c)}_n=
\sum_{k<l=1}^n\sum_{\alpha=1}^2\sum_{\beta=j}^{j+1}
\calc{A}^{(4)}_n(k,l,\alpha,\beta)
\sum_{m=2}^{j-1}
\int d\calc{M}J^{(\alpha,\beta)(b,c)}_{m}(l_1,l_2),
\end{equation}
where
\begin{equation}
J^{(\alpha,\beta)(b)}_{m}(l_1,l_2)
=
\left\{
\begin{split}
 &-\frac{\lr{l_1l_2}\lr{jl_2}}
{\lr{2j}}\lr{2l_1}\calc{D}(2,m), &\textrm{for $\alpha=1$},\\
&\left(-\frac{\left(\lr{2l_1}\lr{\beta l_2}+
\lr{\beta l_1}\lr{2l_2}\right)\lr{\beta l_2}}
{\lr{2\beta}^2}
+\frac{\lr{j\beta}\lr{\beta l_1}\lr{2l_2}^2}
{\lr{2j}\lr{2\beta}^2}
\right)&\hspace{-4mm}\lr{2l_1}\calc{D}(2,m),\\
& &\textrm{for $\alpha=2$},
\end{split} 
\right.
\end{equation}
and
\begin{equation}
J^{(\alpha,\beta)(c)}_{m}(l_1,l_2)
=
\left\{
\begin{split}
&\left(\frac{\left(\lr{jl_1}\lr{\alpha l_2}+
\lr{\alpha l_1}\lr{jl_2}\right)\lr{\alpha l_2}}
{\lr{j\alpha}^2}
-\frac{\lr{2\alpha}\lr{\alpha l_1}\lr{jl_2}^2}
{\lr{j2}\lr{j\alpha}^2}
\right)&\hspace{-4mm}\lr{jl_1}\calc{D}(j+1,m),\\
& &\textrm{for $\beta=j$},\\
 &\frac{\lr{l_1l_2}\lr{2l_2}}
{\lr{j2}}\lr{jl_1}\calc{D}(j+1,m), &\textrm{for $\beta=j+1$}.
\end{split} 
\right.
\end{equation}
The integrand $J^{(\alpha,\beta)(b,c)}_{m}(l_1,l_2)$ 
can be decomposed as
\begin{equation}
J^{(\alpha,\beta)(b,c)}_{m}(l_1,l_2)
=S_{b,c}^{(\alpha,\beta)}(m)-S_{b,c}^{(\alpha,\beta)}(m+1),
\end{equation}
with
\begin{equation}
S_b^{(\alpha,\beta)}(m)=
\left\{
\begin{split}
&
-\frac{\tr(2j1l_1)\tr(j1ml_2)}
{16(1\cdot j)(2\cdot j)(1\cdot l_1)(m\cdot l_2)}
+\frac{\tr(j2ml_2)}{4(2\cdot j)(m\cdot l_2)},
&\textrm{for $\alpha=1$},
\\
&&\\
&
-3\frac{\tr(\beta 21l_1)\tr(\beta 2ml_2)}
{16(2\cdot \beta)^2(1\cdot l_1)(m\cdot l_2)}
+\frac{\tr(\beta 21l_1)\tr(j2ml_2)}
{16(2\cdot j)(2\cdot \beta)(1\cdot l_1)(m\cdot l_2)}\\
&\hspace{1cm}
-\frac{\tr(2\beta 1l_1)\tr(\beta 1ml_2)}
{16(1\cdot\beta)(2\cdot\beta)(1\cdot l_1)(m\cdot l_2)}
+\frac{\tr(\beta 2ml_2)}{4(2\cdot\beta)(m\cdot l_2)},
&\textrm{for $\alpha=2$},
\end{split}
\right. 
\end{equation}
and
\begin{equation}
S_c^{(\alpha,\beta)}(m)=
\left\{
\begin{split}
&
3\frac{\tr(\alpha j(j+1)l_1)\tr(\alpha jml_2)}
{16(\alpha\cdot j)^2(j+1\cdot l_1)(m\cdot l_2)}
-\frac{\tr(\alpha j(j+1)l_1)\tr(2jml_2)}
{16(\alpha\cdot j)(2\cdot j)(j+1\cdot l_1)(m\cdot l_2)}\\
&\hspace{1cm}
+\frac{\tr(j\alpha(j+1)l_1)\tr(\alpha(j+1)ml_2)}
{16(\alpha\cdot j)(\alpha\cdot j+1)(j+1\cdot l_1)(m\cdot l_2)}
-\frac{\tr(\alpha jml_2)}{4(\alpha\cdot j)(m\cdot l_2)},
&\textrm{for $\beta=j$},\\
&&\\
&
\frac{\tr(j2(j+1)l_1)\tr(2(j+1)ml_2)}
{16(2\cdot j)(2\cdot j+1)(j+1\cdot l_1)(m\cdot l_2)}
-\frac{\tr(2jml_2)}{4(2\cdot j)(m\cdot l_2)},
&\textrm{for $\beta=j+1$}.
\end{split}
\right. 
\end{equation}
The loop integral can be easily evaluated similarly
to those in the previous cases.

The diagrams (d) and (e) can be obtained by
rotating the diagrams (b) and (c) upside down,
thus exchanging indices $(1,2)\leftrightarrow (j,j+1), 
l_1\leftrightarrow l_2$ and $m\rightarrow m-1$. 
The index $m$ is then summed over the region $j+2\le m\le n+1$. 

The diagram (f) gives a contribution in the case
$(\alpha,\beta)=(1,j)$, $(2,j+1)$ that is the same
order in $N_c$ as the first five diagrams,
which fill in the final pieces of $\calc{A}^R_n$:
\begin{align}
\calc{A}^{R(f)}_n=&
\sum_{k<l=1}^n\sum_{\alpha=1}^2\sum_{\beta=j}^{j+1}
\calc{A}^{(4)}_n(k,l,\alpha,\beta)
\int d\calc{M}J^{(\alpha,\beta)(f)}_{m}(l_1,l_2),
\nonumber\\
&\hspace{6cm}
\textrm{for $(\alpha,\beta)=(1,j),\ (2,j+1)$},
\end{align}
with
\begin{align}
 J^{(\alpha,\beta)(f)}_{m}(l_1,l_2)=&
-\frac{\lr{\alpha\check{\alpha}}\lr{\beta\check{\beta}}
\left(\lr{\alpha l_1}\lr{\beta l_2}+2\lr{\beta l_1}
\lr{\alpha l_2}\right)}
{\lr{\alpha\beta}^2\lr{\check{\alpha}l_1}\lr{\check{\beta}l_2}}
\nonumber\\
=&
3\frac{\tr(\beta\alpha\check{\alpha}l_1)\tr(\alpha\beta\check{\beta}l_2)}
{16(\alpha\cdot\beta)^2(\check{\alpha}\cdot l_1)
(\check{\beta}\cdot l_2)}
+\frac{\tr(\check{\beta}\alpha\check{\alpha}l_1)
\tr(\beta\alpha\check{\beta}l_2)}{16(\alpha\cdot\beta)
(\alpha\cdot\check{\beta})
(\check{\alpha}\cdot l_1)(\check{\beta}\cdot l_2)}
\nonumber\\
&
-\frac{\tr(\beta\alpha\check{\alpha}l_1)}{4(\alpha\cdot\beta)
(\check{\alpha}\cdot l_1)}.
\end{align}

We need not consider the diagram (g)
since it only contributes to the primitive amplitude 
$\calc{A}_n^{\calc{N}=4}$, not to $\calc{A}_n^R$.

Summing up all the contributions, $\calc{A}^R_n$ for
the four-quark MHV amplitudes are given by
\begin{align}\label{4quarkR}
 \calc{A}^{R}_n=&
\sum_{k<l=1}^n\sum_{\alpha=1}^2\sum_{\beta=j}^{j+1}
c_\Gamma\calc{A}^{(4)}_n(k,l,\alpha,\beta)
\Bigg[
-\frac{1}{\epsilon^2}
\left(\frac{\mu^2}{-s_{12}}\right)^\epsilon
-\frac{1}{\epsilon^2}
\left(\frac{\mu^2}{-s_{jj+1}}\right)^\epsilon
\nonumber\\
&
+{\sum_{m=j+1}^{n+1}\sum_{\substack{a=2\\
\hspace{-14mm}(m,a)\ne(1,2),(j+1,j)}}^j}
b^{\alpha\beta}_{ma}B(m,a)
+\frac{1}{1-2\epsilon}\left(
\sum_{m=j+1}^{n+1}\sum_{a=2}^{j-1}
+\sum_{m=2}^j\sum_{a=j+1}^n
\right)c^{\alpha\beta}_{ma}T_\epsilon(m,a)
\nonumber\\
&
+\frac{1}{1-2\epsilon}
\sum_{m\in\Gamma_1(\alpha,\beta)}\Big(
\tilde{c}^{(1)\alpha\beta}_mT_\epsilon(m,1)
+
d^{(1)\alpha\beta}_mK_\epsilon(m,1)
\Big)
\nonumber\\
&
+\frac{1}{1-2\epsilon}
\sum_{m\in\Gamma_j(\alpha,\beta)}\Big(
\tilde{c}^{(j)\alpha\beta}_m T_\epsilon(m,j)
+d^{(j)\alpha\beta}_mK_\epsilon(m,j)
\Big)
\Bigg],
\end{align}
where the summation regions $\Gamma_1$ and 
$\Gamma_j$ are defined by
\begin{align}
(\Gamma_1(\alpha,\beta),\Gamma_j(\alpha,\beta))=&
\left\{
\begin{matrix}
 (\{j+1,\cdots,n\},\{2,\cdots,j-1\}),
&\textrm{for $(\alpha,\beta)=(1,j)$},\hfill\\
(\{j+2,\cdots,n\},\{j+2,\cdots,n\}),
&\textrm{for $(\alpha,\beta)=(1,j+1)$},\hfill\\
 (\{3,\cdots,j-1\},\{3,\cdots,j-1\}),
&\textrm{for $(\alpha,\beta)=(2,j)$},\hfill\\
(\{3,\cdots,j\},\{j+2,\cdots,n+1\}),
&\textrm{for $(\alpha,\beta)=(2,j+1)$}.\hfill
\end{matrix}
\right. \qquad
\end{align}
We next introduce a combination of the bubble function
$K_\epsilon$ as
\begin{equation}
K_\epsilon(p,P,Q)=\frac{\left(\mu^2\right)^\epsilon}{\epsilon}
\left((-P^2)^{-\epsilon}-(-Q^2)^{-\epsilon}\right),
\label{ebubble}
\end{equation}
and use the same abbreviated notation as in (\ref{triabb}).
For the case $(\alpha,\beta)=(1,j),\ (2,j+1)$,
the coefficient functions are given by
\begin{subequations}
\begin{align}
 b^{\alpha\beta}_{ma}=&
\frac{\tr(\alpha\check{\beta}ma)\tr(\check{\alpha}\beta am)}
{2s_{\alpha\check{\beta}}s_{\check{\alpha}\beta}s_{ma}^2}
-3\frac{\tr(\alpha\beta ma)\tr(\alpha\beta am)}
{2s_{\alpha\beta}^2s_{ma}^2}
\nonumber\\
&
+\frac{\tr(\alpha\beta ma)}{2s_{\alpha\beta}s_{ma}}
\left(
\frac{\tr(\alpha\check{\beta}am)}{s_{\alpha\check{\beta}}s_{ma}}
+
\frac{\tr(\check{\alpha}\beta am)}{s_{\check{\alpha}\beta}s_{ma}}
\right)+(m\leftrightarrow a),\\
c^{\alpha\beta}_{ma}=&\calc{C}(a)-\calc{C}(a+1),\\
\calc{C}(A)=&
-\frac{\tr(\alpha\check{\beta}\check{\alpha}\beta Am)
\tr(\alpha\beta mP_{m+1a})
+\tr(\beta\check{\alpha}\check{\beta}\alpha Am)
\tr(\beta\alpha mP_{m+1a})}
{2s_{\alpha\check{\beta}}s_{\check{\alpha}\beta}s_{\alpha\beta}s_{mA}}
\nonumber\\
&
-3\frac{\tr(\alpha\beta Am)\tr(\alpha\beta mP_{m+1a})
+\tr(\beta\alpha Am)\tr(\beta\alpha mP_{m+1a})}
{2s_{\alpha\beta}^2s_{mA}}
\nonumber\\
&
+\frac{\tr(\alpha\beta Am)\tr(\alpha\check{\beta}mP_{m+1a})
+\tr(\alpha\check{\beta} Am)\tr(\alpha\beta mP_{m+1a})}
{s_{\alpha\beta}s_{\alpha\check{\beta}}s_{mA}}
\nonumber\\
&
+\frac{\tr(\beta\alpha Am)\tr(\beta\check{\alpha}mP_{m+1a})
+\tr(\beta\check{\alpha} Am)\tr(\beta\alpha mP_{m+1a})}
{s_{\alpha\beta}s_{\check{\alpha}\beta}s_{mA}},\\
\tilde{c}^{(1)\alpha\beta}_m=&
\frac{\gamma(\alpha)}{2}\left(
3\frac{\tr(\alpha\beta m\check{\alpha})}
{s_{\alpha\beta}s_{\check{\alpha}m}}
-\frac{\tr(\alpha\check{\beta}m\check{\alpha})}
{s_{\alpha\check{\beta}}s_{\check{\alpha}m}}
\right)
\frac{\tr(\alpha\beta mP_{m+11})-\tr(\beta\alpha mP_{m+11})}
{s_{\alpha\beta}},\\
\tilde{c}^{(j)\alpha\beta}_m=&
\frac{\gamma(\alpha)}{2}\left(
3\frac{\tr(\beta\alpha m\check{\beta})}
{s_{\alpha\beta}s_{\check{\beta}m}}
-\frac{\tr(\beta\check{\alpha}m\check{\beta})}
{s_{\check{\alpha}\beta}s_{\check{\beta}m}}
\right)
\frac{\tr(\beta\alpha mP_{m+1j})-\tr(\alpha\beta mP_{m+1j})}
{s_{\alpha\beta}},\\
d^{(1)\alpha\beta}_m=&
\gamma(\alpha)\frac{\tr(\alpha\check{\beta}m\check{\alpha})}
{s_{\alpha\check{\beta}}s_{\check{\alpha}m}},\\
d^{(j)\alpha\beta}_m=&
\gamma(\alpha)\frac{\tr(\beta\check{\alpha}m\check{\beta})}
{s_{\check{\alpha}\beta}s_{\check{\beta}m}},
\end{align}
\end{subequations}
and for the case $(\alpha,\beta)=(1,j+1),\ (2,j)$ by
\begin{subequations}
\begin{align}
 b^{\alpha\beta}_{ma}=&
\frac{\tr(\alpha\beta ma)\tr(\check{\alpha}\check{\beta}am)}
{s_{\alpha\beta}s_{\check{\alpha}\check{\beta}}s_{ma}^2}
+
\frac{\tr(\check{\alpha}\check{\beta}ma)\tr(\alpha\beta am)}
{s_{\alpha\beta}s_{\check{\alpha}\check{\beta}}s_{ma}^2}
-2\frac{\tr(\alpha\beta ma)\tr(\alpha\beta am)}
{s_{\alpha\beta}^2s_{ma}^2},\\
c^{\alpha\beta}_{ma}=&\calc{C}(a)-\calc{C}(a+1),\\
\calc{C}(A)=&
\frac{\tr(\check{\alpha}\check{\beta}\beta\alpha Am)
\tr(\check{\alpha}\beta mP_{m+1a})
+\tr(\check{\alpha}\beta Am)
\tr(\check{\alpha}\check{\beta}\beta\alpha mP_{m+1a})}
{2s_{\check{\alpha}\beta}s_{\alpha\beta}
s_{\check{\alpha}\check{\beta}}s_{mA}}
\nonumber\\
&
+\frac{\tr(\check{\beta}\check{\alpha}\alpha\beta Am)
\tr(\check{\beta}\alpha mP_{m+1a})+
\tr(\check{\beta}\alpha Am)
\tr(\check{\beta}\check{\alpha}\alpha\beta mP_{m+1a})}
{2s_{\alpha\check{\beta}}s_{\alpha\beta}
s_{\check{\alpha}\check{\beta}}s_{mA}}
\nonumber\\
&
-\frac{\tr(\alpha\beta Am)\tr(\alpha\beta mP_{m+1a})
+\tr(\beta\alpha Am)\tr(\beta\alpha mP_{m+1a})}
{s_{\alpha\beta}^2s_{mA}}
\nonumber\\
&
+\frac{\tr(\alpha\beta Am)\tr(\alpha\check{\beta} mP_{m+1a})
+\tr(\alpha\check{\beta}Am)\tr(\alpha\beta mP_{m+1a})}
{2s_{\alpha\beta}s_{\alpha\check{\beta}}s_{mA}}
\nonumber\\
&
+
\frac{\tr(\beta\alpha Am)\tr(\beta\check{\alpha} mP_{m+1a})
+\tr(\beta\check{\alpha}Am)\tr(\beta\alpha mP_{m+1a})}
{2s_{\check{\alpha}\beta}s_{\alpha\beta}s_{mA}},\\
\tilde{c}^{(1)\alpha,\beta}_m=&
-2\gamma(\alpha)
\frac{\tr(\beta\alpha\check{\alpha}m)\tr(\beta\alpha mP_{m+11})}
{s_{\alpha\beta}^2s_{\check{\alpha}m}},\\
\tilde{c}^{(j)\alpha,\beta}_m=&
2\gamma(\alpha)
\frac{\tr(\alpha\beta\check{\beta}m)\tr(\alpha\beta mP_{m+1j})}
{s_{\alpha\beta}^2s_{\check{\beta}m}},\\
d^{(1)\alpha\beta}_m=&0,\\
d^{(j)\alpha\beta}_m=&0.
\end{align}
\end{subequations}

Finally, the primitive amplitude $\calc{A}^f_n$
comes from the quark chiral-multiplet loop represented
by the last diagram, (h). The amplitudes in the two types 
of helicity configurations are equal and given by
\begin{equation}
\calc{A}^f_n=-\sum_{k<l=1}^n\sum_{\alpha=1}^2\sum_{\beta=j}^{j+1}
\calc{A}^{(4)}_n(k,l,\alpha,\beta)
\sum_{m_1=j+2}^{n+1}\sum_{m_2=2}^{j-1}\int d\calc{M}
J^{(h)}_{(m_1,m_2)}(l_1,l_2),
\end{equation}
where
\begin{equation}
J^{(h)}_{(m_1,m_2)}(l_1,l_2)=\frac{\lr{(m_1-1)m_1}\lr{m_2(m_2+1)}
\lr{1l_1}\lr{jl_2}\lr{j+1l_1}\lr{2l_2}}
{\lr{2j}\lr{1(j+1)}\lr{(m_1-1)l_1}\lr{l_1m_1}\lr{m_2l_2}\lr{l_2(m_2+1)}}.
\end{equation}
This can be similarly written as
\begin{equation}
 J^{(h)}_{(m_1,m_2)}(l_1,l_2)=
S_h(m_1-1,m_2)-S_h(m_1-1,m_2+1)-S_h(m_1,m_2)+S_h(m_1,m_2+1),
\end{equation}
with
\begin{align}
S_h(m_1,m_2)=&
\frac{\lr{m_1(j+1)}\lr{1l_1}\lr{m_22}\lr{jl_2}}
{\lr{1(j+1)}\lr{m_1l_1}\lr{j2}\lr{m_2l_2}}\nonumber\\
=&
\frac{\tr(1(j+1)m_1l_1)\tr(j2m_2l_2)}
{16(1\cdot (j+1))(2\cdot j)(m_1\cdot l_1)(m_2\cdot l_2)},
\end{align}
or, in a form more useful for later use, as
\begin{align}
S_h(m_1,m_2)=&\frac{1}{16(m_1\cdot l_1)(m_2\cdot l_2)}\times
\nonumber\\
\Bigg(
&\Omega
\frac{\tr(j(j+1)m_1l_1)\tr(j2m_2l_2)}
{4(2\cdot j)(j\cdot j+1)}
-
\tilde{\Omega}
\frac{\tr(2(j+1)m_1l_1)\tr(2jm_2l_2)}
{4(2\cdot j)(2\cdot j+1)}
\nonumber\\
+&
\Omega
\frac{\tr(21m_1l_1)\tr(2jm_2l_2)}
{4(1\cdot 2)(2\cdot j)}
-
\tilde{\Omega}
\frac{\tr(j1m_1l_1)\tr(j2m_2l_2)}
{4(1\cdot j)(2\cdot j)}
\nonumber\\
+&
\Omega
\frac{\tr((j+1)1m_1l_1)\tr((j+1)jm_2l_2)}
{4(1\cdot j+1)(j\cdot j+1)}
-
\tilde{\Omega}
\frac{\tr(1(j+1)m_1l_1)\tr(1jm_2l_2)}
{4(1\cdot j)(1\cdot j+1)}
\nonumber\\
+&
\Omega
\frac{\tr(1(j+1)m_1l_1)\tr(12m_2l_2)}
{4(1\cdot 2)(1\cdot j+1)}
-
\tilde{\Omega}
\frac{\tr((j+1)1m_1l_1)\tr((j+1)2m_2l_2)}
{4(1\cdot j+1)(2\cdot j+1)}
\Bigg),
\end{align}
where
\begin{equation}
 \tilde{\Omega}=\frac{\tr(1(j+1)2j)}{4(1\cdot j+1)(2\cdot j)}.
\end{equation}

After the loop integration, we obtain the final result:
\begin{align}\label{4quarkf}
\calc{A}^f_n=& \sum_{k<l=1}^n\sum_{\alpha=1}^2\sum_{\beta=j}^{j+1}
c_\Gamma\calc{A}^{(4)}_n(k,l,\alpha,\beta)
\nonumber\\
&
\times
\Bigg[
\sum_{m=j+2}^n\sum_{a=3}^{j-1}b_{ma}B(m,a)
+
\frac{1}{1-2\epsilon}\left(\sum_{m=3}^{j-1}\sum_{a=j+1}^n
+\sum_{m=j+2}^n\sum_{a=2}^{j-1}
\right)
c_{ma}T_\epsilon(m,a)
\nonumber\\
&
+\frac{1}{1-2\epsilon}\Bigg(
\sum_{m=2}^{j-1}
d^{(n)}_mK_\epsilon(m,n)
+\sum_{m=3}^j
d^{(j+1)}_mK_\epsilon(m,j+1)
\nonumber\\
&\hspace{1.5cm}
+\sum_{m=j+2}^{n+1}
d^{(2)}_mK_\epsilon(m,2)
+\sum_{m=j+1}^n
d^{(j-1)}_mK_\epsilon(m,j-1)
\Bigg)
\Bigg],
\end{align}
where
\begin{subequations}
\begin{align}
b_{ma}=&\frac{\tr(1(j+1)ma)\tr(j2am)
-s_{ma}\tr(2j1(j+1)ma)}{s_{1j+1}s_{2j}s_{ma}^2},\\
c_{ma}=&-\frac{1}{4}
\Bigg[
\left(\frac{\tr(1(j+1)am)}{s_{1j+1}s_{ma}}
-\frac{\tr(1(j+1)(a+1)m)}{s_{1j+1}s_{ma+1}}\right)
\frac{\tr(2jmP_{m+1a})-\tr(j2mP_{m+1a})}{s_{2j}}
\nonumber\\
&
+\left(\frac{\tr(2jam)}{s_{2j}s_{ma}}
-\frac{\tr(2j(a+1)m)}{s_{2j}s_{ma+1}}\right)
\frac{\tr(1(j+1)mP_{m+1a})-\tr((j+1)1mP_{m+1a})}{s_{1j+1}}
\Bigg],\\
d^{(n)}_m=&-\frac{\tr(j21m)}{4s_{2j}s_{1m}},\\
d^{(j+1)}_m=&\frac{\tr(2j(j+1)m)}{4s_{2j}s_{mj+1}},\\
d^{(2)}_m=&\frac{\tr((j+1)12m)}{4s_{1j+1}s_{2m}},\\
d^{(j-1)}_m=&-\frac{\tr(1(j+1)jm)}{s_{1j+1}s_{mj}}.
\end{align}
\end{subequations}

\section{Five-point amplitudes and collinear singularities}\label{fivepoint}

In the previous section, we extended the CSW rule 
to $\calc{N}=1$ SQCD, incorporating massless chiral multiplets, 
and calculated all the one-loop MHV amplitudes including 
external quarks. There is no \textit{a priori} reason, however, 
that this extended CSW rule should generally give the correct result,
since it does not have any string theoretic interpretation. 
For this reason, in this section, we explicitly consider the five-point 
amplitudes and confirm that they exhibit the correct collinear 
behavior.

\subsection{Collinear behavior of the amplitudes}

Let us begin by explaining the general collinear behavior
of the gauge theory (MHV) amplitudes with external quark
chiral multiplets, as well as gluon vector multiplets.
First, consider the $n$-point tree-level amplitudes with 
a fixed color ordering and an arbitrary helicity configuration.
We can easily see that collinear singularities
arise only when neighboring legs, $a$ and $b$, become collinear
and have the form
\begin{equation}
\calc{A}^{\textrm{tree}}_n
\overset{a||b}{\longrightarrow}
\sum_{h=\pm}
\textrm{Split}^{\textrm{tree}}_{-h}
(A^{h_a},B^{h_b})
\calc{A}^{\textrm{tree}}_{n-1}
(\cdots (a+b)^h\cdots), 
\end{equation}
where $A$ and $B$ denote the species of the particles
and the corresponding $h_a$ and $h_b$
represent the signs of their helicities. 
The species of the intermediate state is determined 
by the two collinear particles $A$ and $B$.  
The same symbols $a$ and $b$ appearing in the arguments of 
the amplitudes denote the momenta of 
the corresponding particles.
The non-vanishing splitting amplitudes diverge as
$1/\lr{ab}\sim1/\sqrt{s_{ab}}$ in the collinear limit,
$s_{ab}\rightarrow0$. In this limit, we have $p_a=zP$ and 
$p_b=(1-z)P$, where $P$ and $h$ are the momentum 
and helicity of the intermediate state. 
The tree-level-splitting amplitudes
$\textrm{Split}^{\textrm{tree}}_{-h}(A^{h_a},B^{h_b})$ 
can be found in Ref.~\citen{BDDK4} and references therein.
It should be noted that these splitting amplitudes 
$\textrm{Split}^{\textrm{tree}}_{-h}
(A^{h_a},B^{h_b})$ are universal. They depend only on
two collinear legs but not on the specific amplitude.
We summarize the tree-level splitting amplitudes needed to
investigate the MHV amplitudes in Appendix~\ref{treesplit}.

The collinear limits of the one-loop amplitudes
have the form
\begin{align}
\calc{A}^{\textrm{loop}}_n
\overset{a||b}{\longrightarrow}&
\sum_{\lambda=\pm}\Bigg(
\textrm{Split}^{\textrm{tree}}_{-h}
(A^{h_a},B^{h_b})
\calc{A}^{\textrm{loop}}_{n-1}
(\cdots(a+b)^h\cdots)
\nonumber\\
&\hspace{40mm}
+
\textrm{Split}^{\textrm{loop}}_{-h}
(A^{h_a},B^{h_b})
\calc{A}^{\textrm{tree}}_{n-1}
(\cdots(a+b)^h\cdots)
\Bigg).
\end{align}
Because the one-loop MHV amplitudes of the $\calc{N}=1$ SQCD, obtained
in the previous section, are proportional to the corresponding
tree-level MHV amplitudes, the loop-splitting amplitudes must be 
proportional to the tree-splitting amplitudes,
\begin{equation}
\textrm{Split}^{\textrm{loop}}_{-h}
(A^{h_a},B^{h_b})
=c_\Gamma N_c\times 
\textrm{Split}^{\textrm{tree}}_{-h}
(A^{h_a},B^{h_b})\times
r_S(-h,A^{h_a},B^{h_b}),
\end{equation}
where the extra $N_c$ comes from the one-loop color factor.
Like the amplitudes, the proportionality function 
$r_S(-h,A^{h_a},B^{h_b})$ can be
decomposed into the primitive parts
$r^x_S(-h,A^{h_a},B^{h_b})$
$(x=(\calc{N}=4),(\calc{N}=1 \textrm{chiral}),R,f)$.

\subsection{Five-point amplitudes}\label{fpo}

As an explicit example of our general results, (\ref{2quarkR}), 
(\ref{2quarkf}), (\ref{4quarkR}) and (\ref{4quarkf}), here
we give the five-point amplitudes and investigate 
their collinear behavior to check their consistency. 

The two-quark primitive amplitudes $\calc{A}^R_5$ 
have the form
\begin{equation}
\calc{A}^R_5=\sum_{k<l=1}^5\sum_{\alpha=1}^2\sum_{j=3}^5
c_\Gamma\calc{A}^{(2)}_5(k,l,\alpha,j)
A^R_5(\alpha,j),
\end{equation}
where the quantities $A^R_5(\alpha,j)$ are given as follows,
corresponding to each helicity configuration:
\begin{subequations} 
\begin{align}
A^R_5(1,3)=&-\frac{1}{\epsilon^2}\left(\frac{\mu^2}{-s_{12}}\right)^\epsilon
-\frac{2}{\epsilon}
\left(\frac{\mu^2}{-s_{12}}\right)^\epsilon-4
\nonumber\\
&
+\left(
2\frac{\tr(3124)\tr(3142)}{s_{13}^2}
-\frac{s_{24}\tr(3124)}{s_{13}}
\right)\frac{\textrm{Ls}_1
\left(\frac{-s_{23}}{-s_{51}},\frac{-s_{34}}{-s_{51}}\right)}{s_{15}^2}
\nonumber\\
&
+\left(
2\frac{\tr(3125)\tr(3152)}{s_{13}^2}
-\frac{s_{25}\tr(3125)}{s_{13}}
\right)
\frac{\textrm{Ls}_1\left(
\frac{-s_{51}}{-s_{34}},\frac{-s_{12}}{-s_{34}}\right)}
{s_{34}^2}
\nonumber\\
&
+\frac{\tr(3124)}{s_{13}}\frac{
\textrm{L}_0\left(\frac{-s_{23}}{-s_{51}}\right)
-\textrm{L}_0\left(\frac{-s_{34}}{-s_{51}}\right)}{s_{51}}
+\frac{\tr(3125)}{s_{13}}\frac{
\textrm{L}_0\left(\frac{-s_{51}}{-s_{34}}\right)
-\textrm{L}_0\left(\frac{-s_{12}}{-s_{34}}\right)}{s_{34}},\\
A^R_5(2,3)=&-\frac{1}{\epsilon^2}\left(\frac{\mu^2}{-s_{12}}\right)^\epsilon
-\frac{2}{\epsilon}\left(\frac{\mu^2}{-s_{12}}\right)^\epsilon-4
\nonumber\\
&
-\frac{\tr(3124)}{s_{13}}
\frac{\textrm{Ls}_0\left(\frac{-s_{23}}{-s_{51}},
\frac{-s_{34}}{-s_{51}}\right)}{s_{51}}
-\frac{\tr(3125)}{s_{13}}
\frac{\textrm{Ls}_0\left(\frac{-s_{51}}{-s_{34}},
\frac{-s_{12}}{-s_{34}}\right)}{s_{34}}
\nonumber\\
&
-2\left(
\frac{\tr(3154)}{s_{13}}
-\frac{\tr(3254)}{s_{23}}
\right) \frac{\textrm{L}_0\left(
\frac{-s_{12}}{-s_{34}}\right)}{s_{34}},\\
A^R_5(1,4)=&-\frac{1}{\epsilon^2}\left(\frac{\mu^2}{-s_{12}}\right)^\epsilon
-\frac{2}{\epsilon}\left(\frac{\mu^2}{-s_{12}}\right)^\epsilon-4
\nonumber\\
&
+\left(
2\frac{\tr(4125)\tr(4152)}{s_{14}^2}
-\frac{s_{25}\tr(4125)}{s_{14}}
\right)
\frac{\textrm{Ls}_1\left(\frac{-s_{51}}{-s_{34}},
\frac{-s_{12}}{-s_{34}}\right)}{s_{34}^2}
\nonumber\\
&
+\left(
2\frac{\tr(4135)\tr(4153)}{s_{14}^2}
-\frac{\tr(4135)\tr(4253)}{s_{14}s_{24}}
-\frac{\tr(4235)\tr(4153)}{s_{14}s_{24}}
\right)
\frac{\textrm{Ls}_1\left(\frac{-s_{34}}{-s_{12}},
\frac{-s_{45}}{-s_{12}}\right)}{s_{12}^2}
\nonumber\\
&
-\frac{\tr(4213)}{s_{24}}
\frac{\textrm{Ls}_0\left(\frac{-s_{12}}{-s_{45}},
\frac{-s_{23}}{-s_{45}}\right)}{s_{45}}
+\frac{\tr(4125)}{s_{14}}
\frac{\textrm{L}_0\left(
\frac{-s_{51}}{-s_{34}}\right)
-\textrm{L}_0\left(
\frac{-s_{12}}{-s_{34}}\right)}
{s_{34}}
\nonumber\\
&
+\left(
\frac{\tr(4135)}{s_{14}}
-\frac{\tr(4235)}{s_{24}}
\right)
\frac{\textrm{L}_0
\left(\frac{-s_{34}}{-s_{12}}
\right)
+\textrm{L}_0\left(\frac{-s_{45}}{-s_{12}}
\right)}{s_{12}},\\
A^R_5(2,4)=&-\frac{1}{\epsilon^2}\left(\frac{\mu^2}{-s_{12}}\right)^\epsilon
-\frac{2}{\epsilon}\left(\frac{\mu^2}{-s_{12}}\right)^\epsilon-4
\nonumber\\
&
+\left(
2\frac{\tr(4213)\tr(4231)}{s_{24}^2}
-\frac{s_{13}\tr(4213)}{s_{24}}\right)
\frac{\textrm{Ls}_1\left(
\frac{-s_{12}}{-s_{45}},\frac{-s_{23}}{-s_{45}}\right)}
{s_{45}^2}
\nonumber\\
&
+\left(
2\frac{\tr(4235)\tr(4253)}{s_{24}^2}
-\frac{\tr(4135)\tr(4253)}{s_{14}s_{24}}
-\frac{\tr(4235)\tr(4153)}{s_{14}s_{24}}
\right)
\frac{\textrm{Ls}_1\left(
\frac{-s_{34}}{-s_{12}},\frac{-s_{45}}{-s_{12}}\right)}
{s_{12}^2}
\nonumber\\
&
-\frac{\tr(4125)}{s_{14}}
\frac{\textrm{Ls}_0
\left(\frac{-s_{51}}{-s_{34}},\frac{-s_{12}}{-s_{34}}\right)}
{s_{34}}
-\frac{\tr(4213)}{s_{24}}
\frac{\textrm{L}_0\left(\frac{-s_{12}}{-s_{45}}\right)
-\textrm{L}_0\left(\frac{-s_{23}}{-s_{45}}\right)}
{s_{45}}
\nonumber\\
&
+\left(
\frac{\tr(4135)}{s_{14}}-\frac{\tr(4235)}{s_{24}}\right)
\frac{\textrm{L}_0\left(\frac{-s_{34}}{-s_{12}}\right)
+\textrm{L}_0\left(\frac{-s_{45}}{-s_{12}}\right)}
{s_{12}},\\
A^R_5(1,5)=&-\frac{1}{\epsilon^2}\left(\frac{\mu^2}{-s_{12}}\right)^\epsilon
-\frac{2}{\epsilon}\left(\frac{\mu^2}{-s_{12}}\right)^\epsilon-4
\nonumber\\
&
-\frac{\tr(5213)}{s_{25}}
\frac{\textrm{Ls}_0\left(
\frac{-s_{12}}{-s_{45}},\frac{-s_{23}}{-s_{45}}
\right)}{s_{45}}
-\frac{\tr(5214)}{s_{25}}
\frac{\textrm{Ls}_0\left(
\frac{-s_{45}}{-s_{23}},\frac{-s_{51}}{-s_{23}}
\right)}{s_{23}}
\nonumber\\
&
+2\left(
\frac{\tr(5134)}{s_{15}}-\frac{\tr(5234)}{s_{25}}\right)
\frac{\textrm{L}_0\left(\frac{-s_{45}}{-s_{12}}\right)}
{s_{12}},\\
A^R_5(2,5)=&-\frac{1}{\epsilon^2}\left(\frac{\mu^2}{-s_{12}}\right)^\epsilon
-\frac{2}{\epsilon}\left(\frac{\mu^2}{-s_{12}}\right)^\epsilon-4
\nonumber\\
&
+\left(
2\frac{\tr(5213)\tr(5231)}{s_{25}^2}
-\frac{s_{13}\tr(5213)}{s_{25}}
\right)
\frac{\textrm{Ls}_1\left(
\frac{-s_{12}}{-s_{45}},\frac{-s_{23}}{-s_{45}}\right)}
{s_{45}^2}
\nonumber\\
&
+\left(
2\frac{\tr(5214)\tr(5241)}{s_{25}^2}
-\frac{s_{14}\tr(5214)}{s_{25}}
\right)
\frac{\textrm{Ls}_1\left(
\frac{-s_{45}}{-s_{23}},\frac{-s_{51}}{-s_{23}}\right)}
{s_{23}^2}
\nonumber\\
&
-\frac{\tr(5213)}{s_{25}}
\frac{\textrm{L}_0\left(\frac{-s_{12}}{-s{45}}\right)
-\textrm{L}_0\left(\frac{-s_{23}}{-s_{45}}\right)}
{s_{45}}
-\frac{\tr(5214)}{s_{25}}
\frac{\textrm{L}_0\left(\frac{-s_{45}}{-s_{23}}\right)
-\textrm{L}_0\left(\frac{-s_{51}}{-s_{23}}\right)}
{s_{23}}.
\end{align}
\end{subequations}
The definitions of the functions $\textrm{L}_0$,
$\textrm{Ls}_0$ and $\textrm{Ls}_1$ appearing here 
are summarized in Appendix~\ref{functions}.

The general collinear behavior explained in the previous
subsection requires these functions $A^R_5(\alpha,j)$
to behave as
\begin{equation}
 A^R_5(\alpha,j)
\overset{a||b}{\longrightarrow}
A^R_4+r^R_S(-h,a^{h_a},b^{h_b}),
\end{equation}
where the configuration of $A^R_4$ is not given 
explicitly but fixed by the two collinear legs.\footnote{
The four-point amplitudes needed here are summarized
in Appendix~\ref{fourpoint}.}
It is straightforward to verify that the above five-point 
amplitudes have the correct collinear singularities 
for all the neighboring pairs of external legs.
We obtain the proportionality functions
\begin{subequations}\label{splitR}
\begin{align}
r^R_S(-,g^+,g^+)=&0,\\
r^R_S(+,g^\pm,g^\mp)=&0,\\
r^R_S(\mp,g^+,\bar{q}^\mp)=&F(1-z,s),\\
r^R_S(\mp,q^\mp,g^+)=&F(z,s),\\
r^R_S(+,\bar{q}^\pm,q^\mp)=&
-\frac{1}{\epsilon^2}\left(\frac{\mu^2}{-s}\right)^\epsilon
-\frac{2}{\epsilon}\left(\frac{\mu^2}{-s}\right)^\epsilon-4,
\end{align}
\end{subequations}
where
\begin{equation}
F(z,s)=-\frac{1}{\epsilon^2}\left(\frac{\mu^2}{z(-s)}\right)^\epsilon
+\frac{1}{\epsilon^2}\left(\frac{\mu^2}{-s}\right)^\epsilon
-\textrm{Li}_2(1-z).
\end{equation}

The two-quark primitive amplitudes $\calc{A}^f_5$
are similarly given by
\begin{equation}
\calc{A}^f_5=\sum_{k<l=1}^5\sum_{\alpha=1}^2\sum_{j=3}^5
c_\Gamma\calc{A}^{(2)}_5(k,l,\alpha,j)A^f_5(j),
\end{equation}
where
\begin{subequations} 
\begin{align}
A^f_5(3)=&\frac{\tr(3154)}{s_{13}}
\frac{\textrm{L}_0\left(\frac{-s_{12}}{-s_{34}}\right)}
{s_{34}},\\
A^f_5(4)=&\frac{\tr(4153)\tr(4235)}{s_{14}s_{24}}
\frac{\textrm{Ls}_1\left(
\frac{-s_{34}}{-s_{12}},
\frac{-s_{45}}{-s_{12}}\right)}{s_{12}^2},\\
A^f_5(5)=&
\frac{\tr(5234)}{s_{25}}\frac{\textrm{L}_0
\left(\frac{-s_{12}}{-s_{45}}\right)}{s_{45}}.
\end{align}
\end{subequations}
These amplitudes have the expected collinear singularities,
and we obtain
\begin{subequations}\label{splitf}
\begin{align}
r^f_S(-,a^+,b^+)=&0,\\
r^f_S(+,a^\pm,b^\mp)=&0,\\
r^f_S(+,\bar{q}^\pm,q^\mp)=&
\frac{1}{\epsilon}\left(\frac{\mu^2}{-s}\right)^\epsilon+2,
\end{align}
\end{subequations}
which coincide with the results given in Refs.~\citen{BDK}
and \citen{BDDKq}.

The four-quark primitive amplitudes
\begin{equation}
 \calc{A}^{R}_5=
\sum_{k<l=1}^5\sum_{\alpha=1}^2\sum_{\beta=3}^4
c_\Gamma\calc{A}^{(4)}_5(k,l,\alpha,\beta) 
A^R_5(\alpha,\beta)
\end{equation}
can be computed from the general formula (\ref{4quarkR})
and are given by the explicit forms of the function
$A^R_5(\alpha,\beta)$:
\begin{subequations} 
\begin{align}
A^R_5(1,3)=&
-\frac{1}{\epsilon^2}\left(\frac{\mu^2}{-s_{12}}\right)^\epsilon
-\frac{1}{\epsilon^2}\left(\frac{\mu^2}{-s_{34}}\right)^\epsilon
-\frac{2}{\epsilon}\left(\frac{\mu^2}{-s_{12}}\right)^\epsilon
-\frac{2}{\epsilon}\left(\frac{\mu^2}{-s_{34}}\right)^\epsilon
-8
\nonumber\\
&
+\left(
3\frac{\tr(1342)\tr(1324)}{s_{13}^2}
-\frac{s_{24}\tr(1342)}{s_{13}}
\right)
\frac{\textrm{Ls}_1\left(
\frac{-s_{23}}{-s_{51}},
\frac{-s_{34}}{-s_{51}}\right)}
{s_{51}^2}
\nonumber\\
&
+\left(
3\frac{\tr(1352)\tr(1325)}{s_{13}^2}
-\frac{\tr(1352)\tr(1425)}{s_{13}s_{14}}
-\frac{s_{25}\tr(1452)}{s_{14}}
\right)
\frac{\textrm{Ls}_1\left(
\frac{-s_{51}}{-s_{34}},
\frac{-s_{12}}{-s_{34}}\right)}
{s_{34}^2}
\nonumber\\
&
-\frac{\tr(1435)}{s_{14}}
\frac{\textrm{Ls}_0\left(
\frac{-s_{34}}{-s_{12}},
\frac{-s_{45}}{-s_{12}}\right)}
{s_{12}}
+\frac{\tr(1342)}{s_{13}}
\frac{\textrm{L}_0\left(\frac{-s_{51}}{-s_{23}}\right)}
{s_{23}}
-\frac{\tr(1452)}{s_{14}}
\frac{\textrm{L}_0\left(\frac{-s_{12}}{-s_{34}}\right)}
{s_{34}}
\nonumber\\
&
-\left(s_{12}+3\frac{\tr(1342)}{s_{13}}\right)
\frac{\textrm{L}_0\left(\frac{-s_{34}}{-s_{51}}\right)}
{s_{51}},\\
A^R_5(2,4)=&
-\frac{1}{\epsilon^2}\left(\frac{\mu^2}{-s_{12}}\right)^\epsilon
-\frac{1}{\epsilon^2}\left(\frac{\mu^2}{-s_{34}}\right)^\epsilon
-\frac{2}{\epsilon}\left(\frac{\mu^2}{-s_{12}}\right)^\epsilon
-\frac{2}{\epsilon}\left(\frac{\mu^2}{-s_{34}}\right)^\epsilon
-8
\nonumber\\
&
+\left(
3\frac{\tr(2413)\tr(2431)}{s_{24}^2}
-\frac{s_{13}\tr(2431)}{s_{24}}
\right)
\frac{\textrm{Ls}_1\left(
\frac{-s_{12}}{-s_{45}},
\frac{-s_{23}}{-s_{45}}\right)}
{s_{45}^2}
\nonumber\\
&
+\left(
3\frac{\tr(2453)\tr(2435)}{s_{24}^2}
-\frac{\tr(1453)\tr(2435)}{s_{14}s_{24}}
-\frac{s_{35}\tr(1435)}{s_{14}}
\right)
\frac{\textrm{Ls}_1\left(
\frac{-s_{34}}{-s_{12}},
\frac{-s_{45}}{-s_{12}}\right)}
{s_{12}^2}
\nonumber\\
&
-\frac{\tr(1452)}{s_{14}}
\frac{\textrm{Ls}_0\left(
\frac{-s_{51}}{-s_{34}},
\frac{-s_{12}}{-s_{34}}\right)}
{s_{34}}
+\frac{\tr(2431)}{s_{24}}
\frac{\textrm{L}_0\left(\frac{-s_{23}}{-s_{45}}\right)}
{s_{45}}
-\frac{\tr(1435)}{s_{14}}
\frac{\textrm{L}_0\left(\frac{-s_{34}}{-s_{12}}\right)}
{s_{12}}
\nonumber\\
&
-\left(s_{34}+3\frac{\tr(4213)}{s_{24}}\right)
\frac{\textrm{L}_0\left(\frac{-s_{45}}{-s_{12}}\right)}
{s_{12}},\\
A^R_5(1,4)=&
-\frac{1}{\epsilon^2}\left(\frac{\mu^2}{-s_{12}}\right)^\epsilon
-\frac{1}{\epsilon^2}\left(\frac{\mu^2}{-s_{34}}\right)^\epsilon
-\frac{2}{\epsilon}\left(\frac{\mu^2}{-s_{12}}\right)^\epsilon
-\frac{2}{\epsilon}\left(\frac{\mu^2}{-s_{34}}\right)^\epsilon
-8
\nonumber\\
&
+\left(
2\frac{\tr(1452)\tr(1425)}{s_{14}^2}
-\frac{s_{25}\tr(1452)}{s_{14}}
\right)
\frac{\textrm{Ls}_1\left(\frac{-s_{51}}{-s_{34}},
\frac{-s_{12}}{-s_{34}}\right)}
{s_{34}^2}
\nonumber\\
&
+\left(
2\frac{\tr(1453)\tr(1435)}{s_{14}^2}
-\frac{s_{35}\tr(1435)}{s_{14}}
\right)
\frac{\textrm{Ls}_1\left(\frac{-s_{34}}{-s_{12}},
\frac{-s_{45}}{-s_{12}}\right)}
{s_{12}^2}
\nonumber\\
&
+\frac{\tr(1452)}{s_{14}}
\frac{\textrm{L}_0\left(\frac{-s_{51}}{-s_{34}}\right)
-\textrm{L}_0\left(\frac{-s_{12}}{-s_{34}}\right)}{s_{34}}
-\frac{\tr(1435)}{s_{14}}
\frac{\textrm{L}_0\left(\frac{-s_{34}}{-s_{12}}\right)
-\textrm{L}_0\left(\frac{-s_{45}}{-s_{12}}\right)}{s_{12}},\\
A^R_5(2,3)=&
-\frac{1}{\epsilon^2}\left(\frac{\mu^2}{-s_{12}}\right)^\epsilon
-\frac{1}{\epsilon^2}\left(\frac{\mu^2}{-s_{34}}\right)^\epsilon
-\frac{2}{\epsilon}\left(\frac{\mu^2}{-s_{12}}\right)^\epsilon
-\frac{2}{\epsilon}\left(\frac{\mu^2}{-s_{34}}\right)^\epsilon
-8
\nonumber\\
&
-\frac{\tr(1452)}{s_{14}}
\frac{\textrm{Ls}_0\left(\frac{-s_{51}}{-s_{34}},
\frac{-s_{12}}{-s_{34}}\right)}{s_{34}}
-\frac{\tr(4153)}{s_{14}}
\frac{\textrm{Ls}_0\left(
\frac{-s_{34}}{-s_{12}},\frac{-s_{45}}{-s_{12}}\right)}{s_{12}}
\nonumber\\
&
-\left(\left(
\frac{\tr(1452)+\tr(4153)}{s_{14}}
\right)
-\left(
\frac{\tr(2345)+\tr(3215)}{s_{23}}
\right)
\right)\frac{\textrm{L}_0
\left(\frac{-s_{12}}{-s_{34}}\right)}{s_{34}}.
\end{align}
\end{subequations}
The collinear singularities of these amplitudes
can be confirmed by using the functions appearing in 
(\ref{splitR}).

Finally, the four-quark primitive amplitudes $A^f_5$ are
similarly given by
\begin{equation}
\calc{A}^f_5=
\sum_{k<l=1}^5\sum_{\alpha=1}^2\sum_{\beta=3}^4
c_\Gamma\calc{A}^{(4)}_5(k,l,\alpha,\beta)A^f_5,
\end{equation}
with
\begin{align}
A^f_5=&
\frac{1}{2\epsilon}\left(\frac{\mu^2}{-s_{12}}\right)^\epsilon
+\frac{1}{2\epsilon}\left(\frac{\mu^2}{-s_{34}}\right)^\epsilon
-\frac{1}{2}\log\left(\frac{-s_{12}}{-s_{34}}\right)+2
+\frac{\tr(1452)}{s_{14}}
\frac{\textrm{L}_0\left(\frac{-s_{12}}{-s_{34}}\right)}{s_{34}}.
\end{align}
The collinear behavior can be checked by using (\ref{splitf}).

\section{Summary and discussion}\label{discuss}

In this paper, we have extended the CSW rule to 
the $\calc{N}=1$ SQCD, incorporating massless quark chiral 
multiplets. The MHV vertices have been arranged into
a manifestly supersymmetric form by introducing the fermionic 
variables, $\eta$ and $\chi$ for the vector multiplet and 
$\eta$ and $\rho$ for the chiral multiplet. 
This formulation involves only the physical helicity 
states without auxiliary fields, and therefore 
provides a simple alternative method efficiently compute 
the perturbative SQCD amplitudes, in a manner that preserves 
the manifest supersymmetry. As an application 
of this extended CSW rule, we have calculated all the one-loop 
MHV amplitudes with arbitrary numbers of external legs including 
one or two external quark-antiquark pairs. 
We have confirmed that the five-point amplitudes 
have the correct collinear singularities as a non-trivial 
check of the results,.
We have given explicit forms of (some of) the one-loop splitting 
amplitudes for the $\calc{N}=1$ SQCD. 

As a confirmation of our general expressions, the investigations
of the collinear singularities should be extended to arbitrary 
amplitudes, for which the splitting amplitudes obtained 
in this paper should be useful. 
As another confirmation, it is important 
to extend recently obtained a direct proof which shows 
that the one-loop gluon amplitudes coincide with 
the field theoretical results in $\calc{N}=1$ SYM.\cite{BSTnew} 
We hope to report on progress on these remaining problems 
in the near future.

One of the important applications of 
the CSW rule is computing real perturbative 
QCD amplitudes.\cite{BBST2,BDKquark}
The results obtained in this paper are not directly related 
to QCD but can be used to study several models beyond 
the Standard Model, such as the minimal supersymmetric Standard 
Model or supersymmetric grand unified theories.  
It is expected that the results will be useful for analyzing 
perturbative properties of such models.

It would also be interesting to formulate a twistor-string 
interpretation of the extended CSW rule. Such a twistor string theory, 
which may be related to $\calc{N}=1$ supersymmetric 
gauge theory, is suggested in Ref.~\citen{W} 
as a string on some weighted supertwistor space. 
The fermionic variables $\eta$ and $\chi$ can be interpreted as 
momentum variables of the fermionic coordinates 
in this supertwistor space. However, there are some 
difficulties with this naive expectation,
as discussed in Ref.~\citen{W}. 
We hope that our study provides some insight that may help
to this problem.

\section*{Acknowledgements}

I would like to thank Shigeki Sugimoto and Tatsuya Tokunaga
for discussions and Yosuke Imamura for giving a lecture on twistor 
string theory at YITP. I also would like to thank Hiroyuki Kawamura
for suggesting that I study the collinear behavior of the amplitudes.
This work is supported 
in part by a Grant-in-Aid for Scientific Research (No.13135213) and
by a Grant-in-Aid for the 21st Century COE ``Center for Diversity 
and Universality in Physics'' from the Ministry of Education, 
Culture, Sports, Science and Technology (MEXT) of Japan.

\appendix
\section{Spinor Conventions and Useful Formulas}\label{sconv}

In this paper we use the two-component notation defined by
\begin{align}
\gamma_\mu=&
\begin{pmatrix}
 0&(\sigma_\mu)_{a\dot b} \\
 (\bar\sigma_\mu)^{\dot ab}&0 \\ 
\end{pmatrix},
\end{align}
with
\begin{equation}
(\sigma_\mu)_{a\dot b}=(1_2,\sigma_i),\quad
(\bar\sigma_\mu)^{\dot ab}=(1_2,-\sigma_i),
\end{equation}
where $1_2$ denotes a $2\times2$ unit matrix and
the quantities $\sigma_i$ are
the Pauli matrices. The antisymmetric spinors are defined
by $\epsilon^{01}=\epsilon_{01}=\epsilon^{\dot{0}\dot{1}}
=\epsilon_{\dot{0}\dot{1}}=1$.
We next introduce the shorthand notation
$\lr{ij}$ and $[\bar{i},\bar{j}]$ as
\begin{subequations} 
\begin{align}
\lr{\lambda_i\lambda_j}=&\lambda_{ia}\lambda^a_j
=\epsilon^{ab}\lambda_{ia}\lambda_{jb}
=\epsilon_{ab}\lambda^a_i\lambda^b_j,\\
[\bar{\lambda}_i,\bar{\lambda}_j]=&\bar{\lambda}_{i\dot{a}}
\bar{\lambda}^{\dot{a}}_j
=\epsilon^{\dot{a}\dot{b}}\bar{\lambda}_{i\dot a}\bar{\lambda}_{j\dot{b}}
=\epsilon_{\dot{a}\dot{b}}\bar{\lambda}^{\dot a}_i\bar{\lambda}^{\dot{b}}_j.
\end{align} 
\end{subequations}
Then, with the trace convention
\begin{equation}
\tr(ij\cdots k)={(\sigma^\mu\bar{\sigma}^\nu\cdots\bar{\sigma}^\rho)_a}^a
p_{i\mu}p_{j\nu}\cdots p_{k\rho},
\end{equation}
we can factorize traces contracted with massless momenta as,
for example,
\begin{subequations} 
\begin{align}
\tr(ij)=&
-\lr{ij}\left[\bar{j}\bar{i}\right]
=2(i\cdot j),\\
\tr(ijkl)=&
\lr{kj}\left[\bar{j}\bar{i}\right]
\lr{il}\left[\bar{l}\bar{k}\right],\\
\tr(ijklmn)=&
-\lr{ml}\left[\bar{l}\bar{k}\right]
\lr{kj}\left[\bar{j}\bar{i}\right]
\lr{in}\left[\bar{n}\bar{m}\right].
\end{align}
\end{subequations}
We can also show that the following quantities
are {\it holomorphic}:
\begin{subequations} 
\begin{align}\label{id1}
\frac{\tr(ijkl)}{4(i\cdot j)(k\cdot l)}
=&
 \frac{\lr{kj}\lr{il}}{\lr{ij}\lr{kl}},\\
\frac{\tr(ijklmn)}{8(i\cdot j)(k\cdot l)(m\cdot n)}
=&
\frac{\lr{ml}\lr{kj}\lr{in}}{\lr{kl}\lr{ij}{\lr{mn}}}.
\end{align}
\end{subequations}

The most frequently used identity is Schouten's
identity, given by
\begin{equation}\label{schouten}
 \lr{ij}\lr{kl}+\lr{jk}\lr{il}+\lr{ki}\lr{jl}=0,
\end{equation}
and the identity\footnote{
The special case with $j=k$ for this identity
was studied in Ref.~\citen{BBST}.}
\begin{align}\label{id2}
&2(m_1\cdot m_2)\tr(ijm_1P)\tr(ikm_2P)
-2(m_1\cdot P)\tr(ijm_1m_2)\tr(ikm_2P)
\nonumber\\
&-2(m_2\cdot P)\tr(ijm_1P)\tr(ikm_2m_1)
+P^2\tr(ijm_1m_2)\tr(ikm_2m_1)=0,
\end{align}
where $p_i, p_j$ and $p_k$ are massless momenta, while
$P$ is not necessarily so.

Now, we present further useful identities used in this paper,
in which $p_i,\ p_j\cdots$ are null vectors, but $P$ and $Q$ are not
necessarily so:
\begin{subequations} 
\begin{align}
\tr(ijkl)=&\tr(jilk)=\tr(klij),\\
\tr(i_1\cdots i_{2n+1}\bar{\sigma}^\mu) 
\tr(j_1\cdots j_{2m+1}\bar{\sigma}_\mu)=&
2\tr(i_1\cdots i_{2n+1}j_{2m+1}\cdots j_1),\\
\tr(ijkl)\tr(ilmn)=&2(i\cdot l)\tr(ijklmn),\\
\tr(ijkl)\tr(kjil)=&s_{ij}s_{jk}s_{kl}s_{li},\\ 
\tr(i_1\cdots i_{2n+1}P)\tr(j_1\cdots j_{2m+1}Q)
=&\tr(i_1\cdots i_{2n+1}Q)\tr(j_1\cdots j_{2m+1}P).
\end{align}
\end{subequations}

\section{The Box, the Triangle and the Bubble Functions}\label{functions}

Here we summarize some properties of the box, the triangle and
the bubble functions. The box function $B(s,t,P^2,Q^2)$ defined 
by Eq.~(\ref{box}) can also be written as\cite{BST}
\begin{equation}\label{boxformula}
B(s,t,P^2,Q^2)=\textrm{Li}_2(1-aP^2) 
+\textrm{Li}_2(1-aQ^2) 
-\textrm{Li}_2(1-as) 
-\textrm{Li}_2(1-at),
\end{equation}
which is frequently used in \S\ref{oneloopamp}. 
The $\epsilon$-dependent
triangle function defined by Eq.~(\ref{etriangle})
has three different $\epsilon\rightarrow0$ limits:
\begin{equation}
T_\epsilon(p,P,Q)
\sim
\left\{
\begin{split}
&\frac{\log\left(Q^2/P^2\right)}
{Q^2-P^2}, 
\qquad&\textrm{for $P^2\ne0,\ Q^2\ne0$},\\  
&-\frac{1}{\epsilon}\frac{1}{P^2}
\left(\frac{\mu^2}{-P^2}\right)^\epsilon,
\qquad&\textrm{for $P^2\ne0,\ Q^2=0$},\\
&-\frac{1}{\epsilon}\frac{1}{Q^2}
\left(\frac{\mu^2}{-Q^2}\right)^\epsilon,
\qquad&\textrm{for $P^2=0,\ Q^2\ne0$}.
\end{split}
\right.\label{triangle}
\end{equation}
Here, total momentum conservation, $p+P+Q=0$,
is assumed.

The combination of the bubble 
function defined by Eq.~(\ref{ebubble}) also has 
three different $\epsilon\rightarrow0$ limits:
\begin{equation}
K_\epsilon(p,P,Q)
\sim
\left\{
\begin{split}
&\log\left(Q^2/P^2\right),
\qquad&\textrm{for $P^2\ne0,\ Q^2\ne0$},\\  
&\frac{1}{\epsilon}
\left(\frac{\mu^2}{-P^2}\right)^\epsilon,
\qquad&\textrm{for $P^2\ne0,\ Q^2=0$},\\
&-\frac{1}{\epsilon}
\left(\frac{\mu^2}{-Q^2}\right)^\epsilon,
\qquad&\textrm{for $P^2=0,\ Q^2\ne0$}.
\end{split}
\right.\label{bubble}
\end{equation}

The functions appearing in the five-point amplitudes 
in \S\ref{fpo} are defined by
\begin{subequations} 
\begin{align}
\textrm{L}_0(r)=&\frac{\log(r)}{1-r},\\
\textrm{Ls}_{-1}(r_1,r_2)=&
\textrm{Li}_2(1-r_1)+\textrm{Li}_2(1-r_2)
+\log r_1\log r_2-\frac{\pi^2}{6},\\
\textrm{Ls}_0(r_1,r_2)=&
\frac{\textrm{Ls}_{-1}(r_1,r_2)}{1-r_1-r_2},\\
\textrm{Ls}_1 (r_1,r_2)=&
\frac{1}{1-r_1-r_2}\left(
\textrm{Ls}_0(r_1,r_2)+\textrm{L}_0(r_1)+\textrm{L}_0(r_2)\right).
\end{align}
\end{subequations}
The functions appearing in the general results 
can be written in terms of these functions in the five-point 
amplitudes as follows. 
For the box function $B(s,t,P^2,Q^2)$, 
a kinematical restriction leads
to $P^2=0$ or $Q^2=0$ for the five-point amplitudes. 
Here we consider the case $Q^2=0$,
since two cases are symmetric:
\begin{align}
B(s,t,P^2,0)=&\textrm{Li}_2\left(1-\frac{P^2}{s}\right)
+\textrm{Li}_2\left(1-\frac{P^2}{t}\right)+\frac{\pi^2}{6}
+\frac{1}{2}\log^2\left(\frac{s}{t}\right)
\nonumber\\
=&-\textrm{Li}_2\left(1-\frac{s}{P^2}\right)
-\textrm{Li}_2\left(1-\frac{t}{P^2}\right)
-\log\left(\frac{s}{P^2}\right)\log\left(\frac{t}{P^2}\right)
+\frac{\pi^2}{6}
\nonumber\\
=&-\textrm{Ls}_{-1}\left(\frac{-s}{-P^2},\frac{-t}{-P^2}\right)
\nonumber\\
=&-\left(\frac{-(p+q)^2}{-P^2}\right)
\textrm{Ls}_{0}\left(\frac{-s}{-P^2},\frac{-t}{-P^2}\right).
\end{align}
For triangle functions with $P^2\ne0$, $Q^2\ne0$, we have
\begin{align}
T_\epsilon(p,P,Q)=&
-\frac{\textrm{L}_0\left(\frac{-Q^2}{-P^2}\right)}{P^2}
\nonumber\\
=&-\frac{\textrm{L}_0\left(\frac{-P^2}{-Q^2}\right)}{Q^2}.
\end{align}

We have used the equality $\textrm{Li}_2(1)=\frac{\pi^2}{6}$ and
two formulas for the dilogarithm function
to study the collinear behavior of the five-point amplitudes:
\begin{subequations} 
\begin{align}
&\textrm{Li}_2(1-r)+\textrm{Li}_2(1-r^{-1})=
-\frac{1}{2}\log^2r,\\
&\textrm{Li}_2(r)+\textrm{Li}_2(1-r)=
\frac{\pi^2}{6}-\log r\log(1-r).
\end{align}
\end{subequations}

\section{Phase Space Integrals}\label{psint}

The basic formulas used to evaluate the Lorentz-invariant phase-space
integrals, Passarino-Veltman reduction,\cite{PV} 
are given in Refs.~\citen{BST}, \citen{BBST} and \citen{BBST2}.
Here we summarize some of the results needed for our calculation.

The first formulas are given by
\begin{subequations} 
\begin{align}
\int \lipsP\frac{l_1^\mu}{(m_1\cdot l_1)}=
-m_1^\mu\frac{2\pi\lambda}{2\epsilon(1-2\epsilon)}
\frac{P^2}{(m_1\cdot P)^2}
+P^\mu\frac{2\pi\lambda}{1-2\epsilon}\frac{1}{(m_1\cdot P)},\\
\int \lipsP\frac{l_2^\mu}{(m_2\cdot l_2)}=
-m_2^\mu\frac{2\pi\lambda}{2\epsilon(1-2\epsilon)}
\frac{P^2}{(m_2\cdot P)^2}
+P^\mu\frac{2\pi\lambda}{1-2\epsilon}\frac{1}{(m_2\cdot P)},
\label{onel}
\end{align}
\end{subequations}
where
\begin{align}
\lambda&=\frac{\pi^{\frac{1}{2}-\epsilon}}
{4\Gamma\left(\frac{1}{2}-\epsilon\right)} 
\left(\frac{P^2}{4}\right)^{-\epsilon}.
\end{align}
Introducing the conventional constant
\begin{equation}
c_\Gamma= \frac{1}{(4\pi)^{D/2}}
\frac{\Gamma(1+\epsilon)\Gamma^2(1-\epsilon)}{\Gamma(1-2\epsilon)},
\end{equation}
we obtain the useful relation
\begin{equation}
\frac{\lambda}{(2\pi)^D}=\frac{c_\Gamma}{4\pi^2[\pi\epsilon \csc(\pi\epsilon)]}
(P^2)^{-\epsilon}.
\end{equation}

The second and the most important formulas are obtained 
by calculating
\begin{equation}
 \mathcal{I}^{\mu\nu}:=
\int\lipsP\frac{l_1^\mu l_2^\nu}{(m_1\cdot l_1)(m_2\cdot l_2)},
\end{equation}
which can be expanded as
\begin{align}
\mathcal{I}^{\mu\nu}=&
\eta^{\mu\nu}\mathcal{I}_0
+m_1^\mu m_1^\nu\mathcal{I}_1
+m_2^\mu m_2^\nu\mathcal{I}_2
+P^\mu P^\nu\mathcal{I}_3
+m_1^\mu m_2^\nu\mathcal{I}_4
\nonumber\\
&\hspace{1cm}
+m_2^\mu m_1^\nu\mathcal{I}_5
+m_1^\mu P^\nu\mathcal{I}_6
+P^\mu m_1^\nu\mathcal{I}_7
+m_2^\mu P^\nu\mathcal{I}_8
+P^\mu m_2^\nu\mathcal{I}_9.\label{twol}
\end{align}
The coefficients $\mathcal{I}_i\ (i=0,1,\cdots,9)$ are given by
\begin{subequations} 
\begin{align}
\mathcal{I}_0&=\frac{1}{(m_1\cdot m_2)N}
\Bigg(-2\left[(m_1\cdot m_2)P^2-(m_1\cdot P)(m_2\cdot P)\right]
\tilde{\mathcal{I}}^{(0,0)}
+(m_1\cdot P)N\tilde{\mathcal{I}}^{(1,0)}
\nonumber\\
&\hspace{19mm}
-(m_2\cdot P)N\tilde{\mathcal{I}}^{(0,1)}
-\frac{1}{4}N^2\tilde{\mathcal{I}}^{(1,1)}
-(m_1\cdot P)^2\tilde{\mathcal{I}}^{(1,-1)}
-(m_2\cdot P)^2\tilde{\mathcal{I}}^{(-1,1)}
\Bigg),\\
\mathcal{I}_1&=\frac{1}{(m_1\cdot m_2)^2N^2}
\Bigg(
4(m_2\cdot P)^2\left[
(m_1\cdot m_2)P^2-(m_1\cdot P)(m_2\cdot P)\right]
\tilde{\mathcal{I}}^{(0,0)}
\nonumber\\
&\hspace{19mm}
+(m_2\cdot P)N^2\tilde{\mathcal{I}}^{(1,0)}
+2(m_2\cdot P)^3N\tilde{\mathcal{I}}^{(0,1)}
+\frac{1}{2}(m_2\cdot P)^2N^2\tilde{\mathcal{I}}^{(1,1)}
\nonumber\\
&\hspace{19mm}
+\left[(m_1\cdot m_2)P^2N+2(m_1\cdot P)^2(m_2\cdot P)^2\right]
\tilde{\mathcal{I}}^{(1,-1)}
+2(m_2\cdot P)^4\tilde{\mathcal{I}}^{(-1,1)}
\Bigg),
\end{align}
\begin{align}
\mathcal{I}_2&=\frac{1}{(m_1\cdot m_2)^2N^2}
\Bigg(
4(m_1\cdot P)^2\left[
(m_1\cdot m_2)P^2-(m_1\cdot P)(m_2\cdot P)\right]
\tilde{\mathcal{I}}^{(0,0)}
\nonumber\\
&\hspace{19mm}
-2(m_1\cdot P)^3N\tilde{\mathcal{I}}^{(1,0)}
+\frac{1}{2}(m_1\cdot P)^2N^2\tilde{\mathcal{I}}^{(1,1)}
-(m_1\cdot P)N^2\tilde{\mathcal{I}}^{(0,1)}
\nonumber\\
&\hspace{19mm}
+2(m_1\cdot P)^4\tilde{\mathcal{I}}^{(1,-1)}
+\left[(m_1\cdot m_2)P^2N+2(m_1\cdot P)^2(m_2\cdot P)^2\right]
\tilde{\mathcal{I}}^{(-1,1)}\Bigg),\\
\mathcal{I}_3&=
\frac{1}{N^2}
\Bigg(
2(m_1\cdot m_2)P^2
\tilde{\mathcal{I}}^{(0,0)}
-(m_1\cdot P)N\tilde{\mathcal{I}}^{(1,0)}
\nonumber\\
&\hspace{19mm}
+(m_2\cdot P)N\tilde{\mathcal{I}}^{(0,1)}
+2(m_1\cdot P)^2\tilde{\mathcal{I}}^{(1,-1)}
+2(m_2\cdot P)^2\tilde{\mathcal{I}}^{(-1,1)}
\Bigg),\\
\mathcal{I}_4&=\frac{1}{(m_1\cdot m_2)^2N^2}
\Bigg(
\left[
3(m_1\cdot m_2)P^2N+4(m_1\cdot P)^2(m_2\cdot P)^2
\right]\tilde{\mathcal{I}}^{(0,0)}
\nonumber\\
&\hspace{19mm}
-(m_1\cdot P)\left[
\frac{3}{2}(m_1\cdot m_2)P^2-2(m_1\cdot P)(m_2\cdot P)
\right]N\tilde{\mathcal{I}}^{(1,0)}
\nonumber\\
&\hspace{19mm}
+(m_2\cdot P)\left[
\frac{3}{2}(m_1\cdot m_2)P^2-2(m_1\cdot P)(m_2\cdot P)
\right]N\tilde{\mathcal{I}}^{(0,1)}
+\frac{1}{4}N^3\tilde{\mathcal{I}}^{(1,1)}
\nonumber\\
&\hspace{19mm}
+2(m_1\cdot P)^2\left[(m_1\cdot m_2)P^2
-(m_1\cdot P)(m_2\cdot P)\right]\tilde{\mathcal{I}}^{(1,-1)}
\nonumber\\
&\hspace{19mm}
+2(m_2\cdot P)^2\left[(m_1\cdot m_2)P^2
-(m_1\cdot P)(m_2\cdot P)\right]\tilde{\mathcal{I}}^{(-1,1)}
\Bigg),\\
\mathcal{I}_5&=\frac{1}{(m_1\cdot m_2)^2N^2}
\Bigg(
\left[
3(m_1\cdot m_2)P^2N+4(m_1\cdot P)^2(m_2\cdot P)^2
\right]\tilde{\mathcal{I}}^{(0,0)}
\nonumber\\
&\hspace{19mm}
-(m_1\cdot P)\left[
\frac{3}{2}(m_1\cdot m_2)P^2-2(m_1\cdot P)(m_2\cdot P)
\right]N\tilde{\mathcal{I}}^{(1,0)}
\nonumber\\
&\hspace{19mm}
+(m_2\cdot P)\left[
\frac{3}{2}(m_1\cdot m_2)P^2-2(m_1\cdot P)(m_2\cdot P)
\right]N\tilde{\mathcal{I}}^{(0,1)}
+\frac{1}{4}N^3\tilde{\mathcal{I}}^{(1,1)}
\nonumber\\
&\hspace{19mm}
+2(m_1\cdot P)^2\left[(m_1\cdot m_2)P^2
-(m_1\cdot P)(m_2\cdot P)\right]\tilde{\mathcal{I}}^{(1,-1)}
\nonumber\\
&\hspace{19mm}
+2(m_2\cdot P)^2\left[(m_1\cdot m_2)P^2
-(m_1\cdot P)(m_2\cdot P)\right]\tilde{\mathcal{I}}^{(-1,1)}
\Bigg),\\
\mathcal{I}_6&=\frac{1}{(m_1\cdot m_2)N^2}
\Bigg(
-(m_2\cdot P)\left[
3(m_1\cdot m_2)P^2-2(m_1\cdot P)(m_2\cdot P)\right]
\tilde{\mathcal{I}}^{(0,0)}
\nonumber\\
&\hspace{19mm}
-\frac{1}{2}N^2\tilde{\mathcal{I}}^{(1,0)}
-2(m_2\cdot P)^2N\tilde{\mathcal{I}}^{(0,1)}
-\frac{1}{2}(m_2\cdot P)N^2\tilde{\mathcal{I}}^{(1,1)}
\nonumber\\
&\hspace{19mm}
-(m_1\cdot P)(m_1\cdot m_2)P^2
\tilde{\mathcal{I}}^{(1,-1)}
-2(m_2\cdot P)^3\tilde{\mathcal{I}}^{(-1,1)}
\Bigg),\\
\mathcal{I}_7&=\frac{1}{(m_1\cdot m_2)N^2}
\Bigg(
-(m_2\cdot P)\left[3(m_1\cdot m_2)P^2-2(m_1\cdot P)(m_2\cdot P)
\right]\tilde{\mathcal{I}}^{(0,0)}
\nonumber\\
&\hspace{19mm}
+\frac{1}{2}
(m_1\cdot m_2)P^2
N\tilde{\mathcal{I}}^{(1,0)}
-(m_2\cdot P)^2N\tilde{\mathcal{I}}^{(0,1)}
\nonumber\\
&\hspace{19mm}
-(m_1\cdot P)
(m_1\cdot n_2)P^2
\tilde{\mathcal{I}}^{(1,-1)}
-2(m_2\cdot P)^3\tilde{\mathcal{I}}^{(-1,1)}
\Bigg),\\
\mathcal{I}_8&=\frac{1}{(m_1\cdot m_2)N^2}
\Bigg(
-(m_1\cdot P)\left[
3(m_1\cdot m_2)P^2-2(m_1\cdot P)(m_2\cdot P)\right]
\tilde{\mathcal{I}}^{(0,0)}
\nonumber\\
&\hspace{19mm}
+(m_1\cdot P)^2N\tilde{\mathcal{I}}^{(1,0)}
-\frac{1}{2}
(m_1\cdot m_2)P^2
N\tilde{\mathcal{I}}^{(0,1)}
\nonumber\\
&\hspace{19mm}
-2(m_1\cdot P)^3\tilde{\mathcal{I}}^{(1,-1)}
-(m_2\cdot P)
(m_1\cdot m_2)P^2
\tilde{\mathcal{I}}^{(-1,1)}
\Bigg),\\
\mathcal{I}_9&=\frac{1}{(m_1\cdot m_2)N^2}
\Bigg(
-(m_1\cdot P)\left[
3(m_1\cdot m_2)P^2-2(m_1\cdot P)(m_2\cdot P)
\right]\tilde{\mathcal{I}}^{(0,0)}
\nonumber\\
&\hspace{19mm}
+2(m_1\cdot P)^2N\tilde{\mathcal{I}}^{(1,0)}
+\frac{1}{2}
N^2\tilde{\mathcal{I}}^{(0,1)}
-\frac{1}{2}(m_1\cdot P)N^2\tilde{\mathcal{I}}^{(1,1)}
\nonumber\\
&\hspace{19mm}
-2(m_1\cdot P)^3\tilde{\mathcal{I}}^{(1,-1)}
-(m_2\cdot P)
(m_1\cdot m_2)P^2
\tilde{\mathcal{I}}^{(-1,1)}
\Bigg),
\end{align}
\end{subequations}
where
\begin{equation}
N=(m_1\cdot m_2)P^2-2(m_1\cdot P)(m_2\cdot P), 
\end{equation}
and $\tilde{\mathcal{I}}^{(a,b)}$
are fundamental integrals defined by
\begin{equation}
 \tilde{\mathcal{I}}^{(a,b)}=\int\frac{\lips}
{(m_1\cdot l_1)^a(m_2\cdot l_2)^b}.
\end{equation}
We only need $\tilde{\mathcal{I}}^{(a,b)}$ for
$a,b=0,\pm1$, which can be computed as
\begin{subequations}\label{Tii} 
\begin{align}
\tilde{\mathcal{I}}^{(0,0)}&=
\frac{2\pi\lambda}{1-2\epsilon}\label{fii},\\
\tilde{\mathcal{I}}^{(1,0)}&=
-\frac{1}{\epsilon}\frac{2\pi\lambda}{(m_1\cdot P)},\\
\tilde{\mathcal{I}}^{(0,1)}&=
\frac{1}{\epsilon}\frac{2\pi\lambda}{(m_2\cdot P)},\\
\tilde{\mathcal{I}}^{(1,1)}&=
-\frac{8\pi\lambda}{N}\left(
\frac{1}{\epsilon}+\log\left(1-\frac{(m_1\cdot m_2)P^2}{N}\right)\right),\\
\tilde{\mathcal{I}}^{(1,-1)}&=
-\frac{2\pi\lambda}{(m_1\cdot P)^2}\left(
\frac{N}{2\epsilon}+\frac{1}{1-2\epsilon}
\left[(m_1\cdot m_2)P^2-(m_1\cdot P)(m_2\cdot P)\right]\right),\\
\tilde{\mathcal{I}}^{(-1,1)}&=
-\frac{2\pi\lambda}{(m_2\cdot P)^2}\left(
\frac{N}{2\epsilon}+\frac{1}{1-2\epsilon}
\left[(m_1\cdot m_2)P^2-(m_1\cdot P)(m_2\cdot P)\right]\right).\label{fif}
\end{align}
\end{subequations}
It should be noted that we set $\eta^\mu_\mu=4=D+2\epsilon$,
which is valid for \textit{dimensional reduction}
regularization, which is adopted in this paper,
at the one-loop level. 
The additional contribution of $2\epsilon$ comes form the so-called
$\epsilon$-scalar. 
Using these explicit forms of the fundamental integrals, 
we obtain
\begin{subequations}\label{psintformulas} 
\begin{align}
& \mathcal{I}_0=\frac{2\pi\lambda}{(m_1\cdot m_2)}\log\left(
1-aP^2\right),\\
&\mathcal{I}_3=-\frac{2\pi\lambda}{1-2\epsilon}\frac{2}{N},\\
&\mathcal{I}_7=\frac{2\pi\lambda}{1-2\epsilon}
\frac{P^2}{(m_1\cdot P)N},\\
&\mathcal{I}_8=\frac{2\pi\lambda}{1-2\epsilon}
\frac{P^2}{(m_2\cdot P)N},\\
&\mathcal{I}_5-\frac{P^2}{2(m_1\cdot m_2)}\mathcal{I}_3=
-\frac{2\pi\lambda}{(m_1\cdot m_2)^2}\log\left(1-aP^2\right),\\
&\mathcal{I}_7+\frac{(m_2\cdot P)}{(m_1\cdot m_2)}\mathcal{I}_3=
\frac{2\pi\lambda}{1-2\epsilon}\frac{1}{(m_1\cdot m_2)(m_1\cdot P)},\\
&\mathcal{I}_8+\frac{(m_1\cdot P)}{(m_1\cdot m_2)}\mathcal{I}_3=
\frac{2\pi\lambda}{1-2\epsilon}\frac{1}{(m_1\cdot m_2)(m_2\cdot P)},
\end{align}
\end{subequations}
where
\begin{equation}
 a=\frac{(m_1\cdot m_2)}{N}.
\end{equation}
The one-loop phase space integrals can be calculated using
these relations.

For the special case $P_L=m_1+m_2$ in the gauge $\hat{\eta}=m_1$
or $\hat{\eta}=m_2$, the integral $\calc{I}^{\mu\nu}$
can be expanded as
\begin{equation}
 \mathcal{I}^{\mu\nu}=\eta^{\mu\nu}\tilde{\mathcal{I}}_0
+m^\mu_1m^\nu_1\tilde{\mathcal{I}}_1
+m^\mu_2m^\nu_2\tilde{\mathcal{I}}_2
+m^\mu_1m^\nu_2\tilde{\mathcal{I}}_3
+m^\mu_2m^\nu_1\tilde{\mathcal{I}}_4,\label{onetwo}
\end{equation}
since $P$ is not independent.
The fundamental integrals $\tilde{\mathcal{I}}^{(a,b)}$ can be 
obtained simply by setting $P=m_1+m_2$ in Eq.(\ref{Tii}), 
except for $\tilde{\mathcal{I}}^{(1,1)}$, which diverges, because $N=0$. 
Fortunately, however, we need only $\tilde{\mathcal{I}}_0$ and 
$\tilde{\mathcal{I}}_4$, which can be obtained without using 
$\tilde{\mathcal{I}}^{(1,1)}$ and evaluated as
\begin{subequations} 
\begin{align}
 \tilde{\mathcal{I}}_0&=-\frac{2\pi\lambda}{(m_1\cdot m_2)}
\left(\frac{1}{\epsilon}+\frac{1}{1-2\epsilon}\right)
=(m_1\cdot m_2)\tilde{\mathcal{I}_1}=(m_1\cdot m_2)\tilde{\mathcal{I}_2},\\
\tilde{\mathcal{I}}_4&=\frac{2\pi\lambda}{(m_1\cdot m_2)^2}\frac{1}
{\epsilon(1-2\epsilon)}.
\end{align}
\end{subequations}

\section{Tree-Splitting Amplitudes}\label{treesplit}

Here we present explicit forms of tree-splitting amplitudes, 
which are needed to study the collinear behavior of 
the one-loop MHV amplitudes. They can be obtained 
from the results given in Ref.~\citen{BDDK4} by using the 
supersymmetry.

The tree-level splitting amplitudes considered here
have the form
\begin{equation}
\textrm{Split}^{\textrm{tree}}_{-h}
(A^{h_a},B^{h_b})
=\frac{f_{-h}(A^{h_a},B^{h_b})}
{\sqrt{z(1-z)}\lr{ab}}.
\end{equation}
The functions $f$ for the case in which two collinear
particles $A$ and $B$ are in the vector multiplets
can be obtained from the tree-level
amplitudes (\ref{agtree}) as
\begin{subequations}
 \begin{align}
&f_-(g^{(+)},g^{(+)})=1,  \\
&f_+(g^{(+)},g^{(-)})=(1-z)^2,\\
&f_+(g^{(-)},g^{(+)})=z^2, \\ 
&f_-(g^{(+)},\Lambda^{(+)})=(1-z)^{1/2},\\ 
&f_-(\Lambda^{(+)},g^{(+)})=z^{1/2},  \\
&f_+(g^{(+)},\Lambda^{(-)})=(1-z)^{3/2},\\  
&f_+(\Lambda^{(-)},g^{(+)})=z^{3/2},\\
&f_+(\Lambda^{(+)},\Lambda^{(-)})=z^{1/2}(1-z)^{3/2},\\ 
&f_+(\Lambda^{(-)},\Lambda^{(+)})=-z^{3/2}(1-z)^{1/2}.  
 \end{align}
\end{subequations}
For the other cases, the functions $f$ can be obtained
from the tree-level amplitudes (\ref{2qtree}).
The functions $f$ for the case in which the particles
in the vector and antichiral multiplets become 
collinear are given by
\begin{subequations}
 \begin{align}
&f_-(g^{(+)},\bar{q}^{(+)})=(1-z)^{1/2},  \\
&f_+(g^{(+)},\bar{q}^{(-)})=(1-z)^{3/2},\\  
&f_-(g^{(+)},\bar{\phi}^{(+)})=1-z,\\ 
&f_+(g^{(+)},\bar{\phi}^{(-)})=1-z,\\
&f_-(\Lambda^{(+)},\bar{q}^{(+)})=z^{1/2}(1-z)^{1/2},  \\
&f_+(\Lambda^{(+)},\bar{\phi}^{(-)})=z^{1/2}(1-z).
 \end{align}
\end{subequations}
The functions $f$ for the two collinear particles 
in the chiral and vector multiplets are
\begin{subequations}
 \begin{align}
&f_-(q^{(+)},g^{(+)})=z^{1/2},  \\
&f_+(q^{(-)},g^{(+)})=z^{3/2},\\  
&f_-(\phi^{(+)},g^{(+)})=z,  \\
&f_+(\phi^{(-)},g^{(+)})=z,\\
&f_-(q^{(+)},\Lambda^{(+)})=z^{1/2}(1-z)^{1/2},\\ 
&f_+(\phi^{(-)},\Lambda^{(+)})=-z(1-z)^{1/2}.
 \end{align}
\end{subequations}
The collinear behavior of the antichiral and 
chiral multiplets is characterized by
\begin{subequations}
 \begin{align}
&f_+(\bar{q}^{(+)},q^{(-)})=z^{1/2}(1-z)^{3/2},\\
&f_+(\bar{q}^{(-)},q^{(+)})=-z^{3/2}(1-z)^{1/2}, \\ 
&f_+(\bar{q}^{(+)},\phi^{(-)})=z^{1/2}(1-z),\\ 
&f_+(\bar{\phi}^{(-)},q^{(+)})=-z(1-z)^{1/2},  \\
&f_+(\bar{\phi}^{(+)},\phi^{(-)})=z(1-z),\\  
&f_+(\bar{\phi}^{(-)},\phi^{(+)})=z(1-z).
 \end{align}
\end{subequations}
We can confirm their universality by using 
the amplitudes (\ref{4qtree}).

\section{Four-Point Amplitudes}\label{fourpoint}

Here we present the four-point amplitudes
needed to study the collinear behavior of the five-point
amplitudes. First, the four-gluon primitive amplitudes
$\calc{A}^{\calc{N}=1 \textrm{chiral}}_4=
-\calc{A}^f_4$ are given by
\begin{align}
\calc{A}^{\calc{N}=1 \textrm{chiral}}_4&=
-\sum_{k<l}^4\sum_{i<j}^4
c_\Gamma\calc{A}^{(0)}_4(k,l,i,j)A^f_4(i,j),
\end{align}
where $A^f_4(i,j)$ is given by
\begin{subequations} 
\begin{align}
A^f_4(1,2)=&-\frac{1}{\epsilon}
\left(\frac{\mu^2}{-s_{23}}\right)^\epsilon-2,\\ 
A^f_4(2,3)=&-\frac{1}{\epsilon}
\left(\frac{\mu^2}{-s_{12}}\right)^\epsilon-2,\\ 
A^f_4(3,4)=&-\frac{1}{\epsilon}
\left(\frac{\mu^2}{-s_{23}}\right)^\epsilon-2,\\ 
A^f_4(4,1)=&-\frac{1}{\epsilon}
\left(\frac{\mu^2}{-s_{12}}\right)^\epsilon-2,\\
A^f_4(1,3)=&-\frac{1}{\epsilon}
\left(\frac{\mu^2}{-s_{12}}\right)^\epsilon-2
\nonumber\\
&\hspace{10mm}
-\frac{s_{12}s_{23}}{2s_{13}^2}\left(
\log^2\left(\frac{-s_{12}}{-s_{23}}\right)+\pi^2\right)
+\frac{s_{12}}{s_{13}}\log\left(\frac{-s_{12}}{-s_{23}}\right),\\ 
A^f_4(2,4)=&-\frac{1}{\epsilon}
\left(\frac{\mu^2}{-s_{23}}\right)^\epsilon-2
\nonumber\\
&\hspace{10mm}
-\frac{s_{12}s_{23}}{2s_{13}^2}\left(
\log^2\left(\frac{-s_{12}}{-s_{23}}\right)+\pi^2\right)
-\frac{s_{12}}{s_{13}}\log\left(\frac{-s_{12}}{-s_{23}}\right).
\end{align}
\end{subequations}

The primitive two-quark primitive amplitudes $\calc{A}^R_4$
have the form
\begin{equation}
\calc{A}^R_4=\sum_{k<l=1}^4\sum_{\alpha=1}^2\sum_{j=3}^4
c_\Gamma\calc{A}^{(2)}_4(k,l,\alpha,j)
A^R_4(\alpha,j),
\end{equation}
where
\begin{subequations} 
\begin{align}
A^R_4(1,3)=&
-\frac{1}{\epsilon^2}\left(\frac{\mu^2}{-s_{12}}\right)^\epsilon
-\frac{2}{\epsilon}\left(\frac{\mu^2}{-s_{12}}\right)^\epsilon-4
\nonumber\\
&\hspace{10mm}
 +\frac{s_{12}(s_{12}-s_{23})}{2s_{13}^2}\left(
\log^2\left(\frac{-s_{12}}{-s_{23}}\right)+\pi^2\right)
+2\frac{s_{12}}{s_{13}}\log\left(\frac{-s_{12}}{-s_{23}}\right),\\
A^R_4(2,3)=&
-\frac{1}{\epsilon^2}\left(\frac{\mu^2}{-s_{12}}\right)^\epsilon
-\frac{2}{\epsilon}\left(\frac{\mu^2}{-s_{12}}\right)^\epsilon-4
\nonumber\\
&\hspace{20mm}
-\frac{s_{12}}{2s_{14}}\left(
\log^2\left(\frac{-s_{12}}{-s_{23}}\right)+\pi^2\right),\\
A^R_4(1,4)=&
-\frac{1}{\epsilon^2}\left(\frac{\mu^2}{-s_{12}}\right)^\epsilon
-\frac{2}{\epsilon}\left(\frac{\mu^2}{-s_{12}}\right)^\epsilon-4
\nonumber\\
&\hspace{20mm}
-\frac{s_{12}}{2s_{14}}\left(
\log^2\left(\frac{-s_{12}}{-s_{23}}\right)+\pi^2\right),\\
A^R_4(2,4)=&
-\frac{1}{\epsilon^2}\left(\frac{\mu^2}{-s_{12}}\right)^\epsilon
-\frac{2}{\epsilon}\left(\frac{\mu^2}{-s_{12}}\right)^\epsilon-4
\nonumber\\
&\hspace{10mm}
 +\frac{s_{12}(s_{12}-s_{23})}{2s_{13}^2}\left(
\log^2\left(\frac{-s_{12}}{-s_{23}}\right)+\pi^2\right)
+2\frac{s_{12}}{s_{13}}\log\left(\frac{-s_{12}}{-s_{23}}\right).
\end{align}
\end{subequations}

The two-quark primitive amplitude 
$\calc{A}^f_4$ is similarly obtained as
\begin{equation}
\calc{A}^f_4=\sum_{k<l=1}^4\sum_{\alpha=1}^2\sum_{j=3}^4
c_\Gamma\calc{A}^{(2)}_4(k,l,\alpha,j)A^f_4(j),
\end{equation}
but we have
\begin{equation}
A^f_4(i,j)=0.
\end{equation}

The four-quark primitive amplitudes $\calc{A}^R_4$ are
give by
\begin{equation}
 \calc{A}^{R}_4=
\sum_{k<l=1}^4\sum_{\alpha=1}^2\sum_{\beta=3}^4
c_\Gamma\calc{A}^{(4)}_4(k,l,\alpha,\beta) 
A^R_4(\alpha,\beta),
\end{equation}
with
\begin{subequations} 
\begin{align}
 A^R_4(1,3)=&-\frac{2}{\epsilon^2}
\left(\frac{\mu^2}{-s_{12}}\right)^\epsilon
-\frac{4}{\epsilon}
\left(\frac{\mu^2}{-s_{12}}\right)^\epsilon-8
\nonumber\\
&\hspace{10mm}
+\frac{s_{12}(s_{12}-2s_{23})}{2s_{13}^2}
\left(\log^2\left(\frac{-s_{12}}{-s_{23}}\right)+\pi^2\right)
+3\frac{s_{12}}{s_{13}}\log\left(\frac{-s_{12}}{-s_{23}}\right),\\
 A^R_4(2,4)=&-\frac{2}{\epsilon^2}
\left(\frac{\mu^2}{-s_{12}}\right)^\epsilon
-\frac{4}{\epsilon}
\left(\frac{\mu^2}{-s_{12}}\right)^\epsilon-8
\nonumber\\
&\hspace{10mm}
+\frac{s_{12}(s_{12}-2s_{23})}{2s_{13}^2}
\left(\log^2\left(\frac{-s_{12}}{-s_{23}}\right)+\pi^2\right)
+3\frac{s_{12}}{s_{13}}\log\left(\frac{-s_{12}}{-s_{23}}\right),\\
A^R_4(1,3)=&-\frac{2}{\epsilon^2}
\left(\frac{\mu^2}{-s_{12}}\right)^\epsilon
-\frac{4}{\epsilon}
\left(\frac{\mu^2}{-s_{12}}\right)^\epsilon-8,\\
A^R_4(2,3)=&
-\frac{2}{\epsilon^2}
\left(\frac{\mu^2}{-s_{12}}\right)^\epsilon
-\frac{4}{\epsilon}
\left(\frac{\mu^2}{-s_{12}}\right)^\epsilon-8.
\end{align}
\end{subequations}
The remaining primitive amplitude $\calc{A}^f_4$ is
\begin{equation}
\calc{A}^f_4=
\sum_{k<l=1}^4\sum_{\alpha=1}^2\sum_{\beta=3}^4
c_\Gamma\calc{A}^{(4)}_4(k,l,\alpha,\beta)A^f_4,
\end{equation}
with
\begin{equation}
A^f_4=\frac{1}{\epsilon}
\left(\frac{\mu^2}{-s_{12}}\right)^\epsilon+2.
\end{equation}


\begin{thebibliography}{99}

\bibitem{W}
E.~Witten, 
\CMP{252,2004,189}; hep-th/0312171.
\bibitem{CSW}
F.~Cachazo, P.~Svrcek and E.~Witten,
\JHEP{09,2004,006}; hep-th/0403047.
\bibitem{GK}
G.~Georgiou and V.~V.~Khoze,
\JHEP{05,2004,070}; hep-th/0404072.
\bibitem{GGK}
G.~Georgiou, E.~W.~N.~Glover and V.~V.~Khoze,
\JHEP{07,2004,048}; hep-th/0407027.
\bibitem{BGK}
S.~D.~Badger, E.~W.~Glover and V.~Khoze,
\JHEP{03,2005,023}; hep-th/0412275.
\bibitem{WZ}
J.-B.~Wu and C.-J.~Zhu,
\JHEP{07,2004,032}; hep-th/0406085.
\bibitem{BST}
A.~Brandhuber, B.~Spence and G.~Travaglini,
\NPB{706,2005,150}, hep-th/0407214.
\bibitem{BBST}
J.~Bedford, A.~Brandhuber, B.~Spence and G.~Travaglini,
\NPB{706,2005,100}; hep-th/0410280.
\bibitem{QR}
C.~Quigley and M.~Rozali,
\JHEP{01,2005,053}; hep-th/0410278.
\bibitem{BDDK4}
Z.~Bern, L.~J.~Dixon, D.~C.~Dunbar and D.~A.~Kosower,
\NPB{425,1994,217}; hep-ph/9403226.
\bibitem{BDDK1}
Z.~Bern, L.~J.~Dixon, D.~C.~Dunbar and D.~A.~Kosower,
\NPB{435,1995,59}; hep-ph/9409265.
\bibitem{WZ2}
J.-B.~Wu and C.-J.~Zhu,
\JHEP{09,2004,063}; hep-th/0406146.
\bibitem{SW}
X.~Su and J.-B.~Wu,
\JL{Mod. Phys. Lett. A,20,2005,1065}; hep-th/0409228.
\bibitem{PR}
J.~Park and S.-J.~Rey,
\JHEP{12,2004,017}; hep-th/0411123.
\bibitem{GKRRTZ}
S.~Giombi, M.~Kulaxizi, R.~Ricci, D.~Robles-Llana, D.~Trancanelli 
and K.~Zoubos,
\NPB{719,2005,234}; hep-th/0411171.
\bibitem{PRep}
M.~L.~Mangano and S.~J.~Parke,
\PRP{200,1991,301}; hep-th/0509223.
\bibitem{BBST2}
J.~Bedford, A.~Brandhuber, B.~Spence and G.~Travaglini,
\NPB{712,2005,59}; hep-th/0412108.
\bibitem{BSTnew}
A.~Brandhuber, B.~Spence and G.~Travaglini,
\textsl{\lq\lq From Trees to Loops and Back''},
QMUL-PH-05-12; hep-th/0510253.
\bibitem{BK}
Z.~Bern and D.A.~Kosower,
\NPB{379,1992,451}.
\bibitem{dilog}
L.~Lewin, \textsl{Dilogarithms and Associated Functions}
(Macdonald, 1958).
\bibitem{BDK}
Z.~Bern, L.~J.~Dixon and D.~A.~Kosower,
\NPB{437,1995,259}; hep-ph/9409393.
\bibitem{BDDKq}
Z.~Bern, L.~Dixon, D.~C.~Dunbar and D.~A.~Kosower,
\textsl{
\lq\lq One Loop Gauge Theory Amplitudes with
an Arbitrary Number of External	Legs''},
\textsl{Proceedings of the Conference on
Continuous Advances in QCD: Theoretical Physics Institute,
University of Minnesota, Minneapolis, USA, 18--20 February 1994},
World Scientific, Singapore 1994; hep-ph/9405248.
\bibitem{BDKquark}
Z.~Bern, L.~J.~Dixon and D.~A.~Kosower,
\textsl{\lq\lq The Last of the Finite Loop Amplitudes in QCD''},
UCLA/05/TEP/15, SLAC-PUB-11134, Saclay/SPhT-T05/058; hep-ph/0505055.
\bibitem{PV}
G.~Passarino and M.~J.~G.~Veltman,
\NPB{160,1979,151}.

\end{thebibliography}
\end{document}